\documentclass{aa}  

\usepackage{graphicx}
\usepackage{soul}        % Highlighting and strikeout

\usepackage{txfonts}
\usepackage{hyperref}
\hypersetup{
    colorlinks=true,
    linkcolor=blue,
    filecolor=blue,      
    urlcolor=blue,
    citecolor=blue,
    pdftitle={Overleaf Example},
    pdfpagemode=FullScreen,
    }% To add links in your PDF file, use the package "hyperref"
\usepackage[dvipsnames]{xcolor}

\newcommand{\vlos}[1][]{v_{los}}
\newcommand{\slos}[1][]{\sigma_{los}}
\newcommand{\kms}[1][]{km s$^{-1}$\kern0pt}
\newcommand{\met}[1][]{{\rm [Fe/H]}}

\newcommand{\PP}{\mathcal{P}}
\newcommand{\LL}{\mathcal{L}}
\newcommand{\MM}{\mathcal{M}}
\newcommand{\VV}{\mathcal{V}}
\newcommand{\EE}{\mathcal{E}}

\newcommand{\Mi}{M_p}

\newcommand{\fpop}{f_p}

\newcommand{\foi}{f_{0,p}}
\newcommand{\Joi}{J_p}
\newcommand{\JMP}{J_{{\rm MP}}}
\newcommand{\JMR}{J_{{\rm MR}}}
\newcommand{\hi}{h_p}
\newcommand{\hMP}{h_{\rm MP}}
\newcommand{\hMR}{h_{\rm MR}}
\newcommand{\hri}{h_{r,p}}

\newcommand{\hzi}{h_{z,p}}
\newcommand{\hzMP}{h_{z,{\rm MP}}}
\newcommand{\hzMR}{h_{z,{\rm MR}}}
\newcommand{\hphii}{h_{\phi,p}}
\newcommand{\gi}{g_p}
\newcommand{\gMP}{g_{\rm MP}}
\newcommand{\gMR}{g_{\rm MR}}
\newcommand{\gri}{g_{r,p}}

\newcommand{\gzi}{g_{z,p}}
\newcommand{\gzMP}{g_{z,{\rm MP}}}
\newcommand{\gzMR}{g_{z,{\rm MR}}}
\newcommand{\gphii}{g_{\phi,p}}
\newcommand{\Bi}{B_p}
\newcommand{\BMP}{B_{\rm MP}}
\newcommand{\BMR}{B_{\rm MR}}
\newcommand{\Gammai}{\Gamma_p}
\newcommand{\GammaMP}{\Gamma_{\rm MP}}
\newcommand{\GammaMR}{\Gamma_{\rm MR}}
\newcommand{\MDM}{M_{\rm DM}}
\newcommand{\rs}{r_s}
\newcommand{\fMP}{f_{\rm MP}}
\newcommand{\fMR}{f_{\rm MR}}
\newcommand{\rcut}{r_{\rm cut}}

\newcommand{\nstar}{n_{\rm star}}
\newcommand{\btheta}{\boldsymbol{\theta}}

\newcommand{\MP}{{\rm MP}}
\newcommand{\MR}{{\rm MR}}
\newcommand{\C}{{\rm C}}
\newcommand{\Rh}{R_h}

\newcommand{\JJ}{\boldsymbol{J}}
\newcommand{\bzeta}{\boldsymbol{\zeta}}

\newcommand{\dd}{{\text{d}}}

\defcitealias{Battaglia2008}{B08}
\defcitealias{Walker2011}{W11}
\defcitealias{Amorisco2012}{AE12}
\defcitealias{Breddels2013a}{B13a}
\defcitealias{Breddels2013b}{B13b}
\defcitealias{Richardson2014}{R14}
\defcitealias{Zhu2016}{Z16}
\defcitealias{Strigari2017}{S17}
\defcitealias{Kaplinghat2019}{K19}
\defcitealias{Read2019}{R19}
\defcitealias{exposito-marquez2023}{E23}
\defcitealias{Hayashi2020}{H20}
\defcitealias{Pascalethesis}{P20}
\defcitealias{ArroyoPolonio2024}{AP24}

\begin{document}

   \title{Chemo-dynamics of the stellar component of the Sculptor dwarf galaxy II: dynamical properties and dark matter halo density.}

   \author{
    José María Arroyo-Polonio \inst{1, 2}\thanks{E-mail: jmarroyo@iac.es (IAC)},
    Raffaele Pascale\inst{3}, 
    Giuseppina Battaglia\inst{1, 2}, \\ 
    Guillaume F. Thomas\inst{1, 2},  
    Carlo Nipoti\inst{4},
    Eugene Vasiliev\inst{5} and
    Eline Tolstoy\inst{6}.
    }

   \institute{Instituto de Astrofísica de Canarias, Calle Vía Láctea s/n E-38206 La Laguna, Santa Cruz de Tenerife, España.    
         \and
             Universidad de La Laguna, Avda. Astrofísico Francisco Sánchez E-38205 La Laguna, Santa Cruz de Tenerife, España.
                     \and
            INAF – Osservatorio di Astrofisica e Scienza dello Spazio di Bologna, Via Piero Gobetti 93/3, 40129 Bologna, Italy
        \and
            Dipartimento di Fisica e Astronomia “Augusto Righi” – DIFA, Alma Mater Studiorum – Universita` di Bologna, via Gobetti 93/2, I-40129 Bologna, Italy.
        \and
        University of Surrey, Guildford, GU2 7XH, UK
        \and
        Kapteyn Astronomical Institute, University of Groningen, PO Box 800, 9700AV Groningen, the Netherlands.
             }

   \date{Submitted on \today; received -; accepted -}

% \abstract{}{}{}{}{} 
% 5 {} token are mandatory
 
  \abstract
   {Dwarf galaxies satellite of the Milky Way are excellent laboratories for testing dark matter (DM) models and baryonic feedback implementation in simulations. The Sculptor "classical" dwarf spheroidal galaxy, a system with two distinct stellar populations and high-quality data, offers a remarkable opportunity to study DM distributions in these galaxies. However, inferences from dynamical modeling in the literature lead to discrepant results. In this work, we infer the DM halo density distribution of Sculptor, applying a method based on spherically symmetric distribution functions depending on actions to fit the stellar structural and kinematic properties of Sculptor. The galaxy is represented via four components: two distinct stellar populations based on distribution functions, tracers within a fixed and dominant DM potential, plus the contribution of a third stellar component that accounts for possible sources of contamination. The model-data comparison accounts for the kinematics and metallicities of individual stars rather than relying on binned profiles, allowing us to assign probabilities of membership to each star. This is the most general approach employed to date to model Sculptor and it is applied on the largest available set of spectroscopic data, which have not been previously analyzed with this objective. We find the DM distribution of Sculptor to have a logarithmic inner slope of $\gamma = 0.39 ^{+0.23}_{-0.26}$ and a scale radius of $r_s = 0.79^{+0.38}_{-0.17}$ kpc at 1-$\sigma$ confidence level. Our results show that Sculptor DM density profile deviates from predictions of DM-only simulations at a 3-$\sigma$ level over a large range of radii. The dynamical-to-luminous mass ratio is around 13 at the 3D half-light radius and 154 at 2 kpc, the outermost radius with observed stars in our dataset. Our analysis suggests that the velocity distribution of Sculptor’s two main stellar components is isotropic in the center and becomes radially anisotropic in the outskirts. Additionally, we provide predictions for the projected radial and tangential velocity dispersion profiles. We also present updated DM annihilation and decay $J$- and $D$-factors, finding $J = 18.15^{+0.11}_{-0.12}$ and $D = 18.07^{+0.10}_{-0.10}$ for an angular aperture of 0.5 degrees.}


   \authorrunning{Arroyo-Polonio J.-M. et al.}
   \titlerunning{Dark matter density profile of the Sculptor dSph}
   \maketitle 
%
%-------------------------------------------------------------------

\section{Introduction} \label{sec:Introduction}

In the current $\Lambda$ cold dark matter ($\Lambda$CDM) cosmological model, pure $N$-body cosmological simulations predict that dark matter (DM) forms halos characterized by cuspy \citet*[NFW]{Navarro1997} density profiles. However, this prediction stands in contrast to a wealth of observational evidence indicating the presence of constant-density cores in the central regions of many low surface brightness galaxies \citep[][among others]{Gentile2004,Simon2005,Deblok2008,Walker2011,Pascale2018,Leung2021}. This is known as the cusp/core problem. To address this tension, as well as others not discussed in this article, modifications to the $\Lambda$CDM model have been proposed, including self-interacting DM (SIDM) \citep{Carlson1992}, fuzzy DM \citep{Hu2000} and others \citep[see][and references therein]{Bullock2017}. Real galaxies, however, are not merely collections of collisionless DM particles, as they also contain stars, dust and gas. In the past years, two main  mechanisms have been identified through which the baryonic component can influence the distribution of DM in the inner regions of the halo: via violent and rapid variations of the gravitational potential due to gas expulsion as a consequence of supernova feedback (e.g. \citealt{Mashchenko2006} \& \citealt{Pontzen2014}), or via dynamical friction with fragmented gas clouds even before stellar feedback takes place (e.g. \citealt{ElZant2001} \& \citealt{Nipoti2015}). Dynamical friction between massive gas clouds and DM is efficient enough to form cores in galaxies with gas mass $M_{gas} = 10^7 M_{\odot}$ and DM halo virial mass $M_{halo} = 10^9 M_{\odot}$, i.e. in the regime of dSphs, as shown by adiabatic, non-cosmological simulations \citep{Nipoti2015}. As for the impact of stellar feedback, typically, some degree of core formation occurs in those simulations that resolve and model star formation in high-density gas, like NIHAO \citep{Wang2015} and FIRE \citep{Hopkins2014}, with the process being particularly efficient in galaxies where the ratio of stellar-to-halo mass is 10$^{-3} < M_{\star}/M_{halo} < 10^{-2}$ \citep{Dicintio2014, onorbe2015}. However, the scales of the smallest and most DM dominated galaxies, like "classical" dwarf spheroidals (dSphs) and ultra faint dwarfs (UFDs), are particularly tricky as this regime of masses and sizes approaches the mass and spatial resolution limits of simulations \citep[see discussion by][]{Sales2022}. In fact, higher-resolution non-cosmological hydrodynamical simulations are able to form cores in the regime of faint galaxies down to $M_{\star}/M_{halo} \approx 10^{-4}$ at $\log\left(M_{\star}[M_{\odot}]\right) = 4$ \citep{Read2016, Pascale2023}, but only in systems with bursty and prolonged star formation histories (SFHs) up to the present. Hence several aspects remain debated, such as the expected core sizes, the minimum mass to form a core and how core formation relates to the galaxy's SFH. In summary, while the inclusion of baryons in the simulations offers a promising solution to the cusp/core problem, it remains unclear whether they can fully reconcile all observations with the $\Lambda$CDM model predictions.  Therefore, accurate and precise inferences of the DM density profiles of galaxies over a range of stellar masses and star formation SFHs are clearly essential to effectively constrain these models.

Local Group (LG) dwarf galaxies are thus precious laboratories in this context.  These galaxies lie in the $M_{\star}/M_{halo}$ regime \citep[][and references therein]{Battaglia2022b} in which the efficacy of stellar feedback in turning cusps into cores varies from extremely low to strong, depending on the exact correspondence between $M_{\star}$ and $M_{halo}$. Furthermore, they display a variety of SFHs \citep{Gallart2015}, in particular when considering M31 and isolated Local Group dwarfs. 
The classical dSphs satellites of the Milky Way are the best studied among the LG dwarf galaxies, due to the amount of observational data that can be gathered for these systems with current facilities. However, they lack gas, leaving stars as the only tracers of the gravitational potential. This poses a significant complication, as the anisotropy of stellar motions is degenerate with the underlying mass distribution. Nevertheless, these systems have been extensively studied over the years, particularly Draco, Sculptor, and Fornax, thanks to the availability of rich spectroscopic datasets. Previous studies have typically found cores in Fornax \citep{Walker2011,Pascale2018,Read2019} and cusps in Draco  \citep{Read2019, Hayashi2020, Vitral2024}, but with large uncertainties and scatter among the results. In the context of feedback-induced modifications to the DM halo properties, this might be considered in line with the lower stellar mass $M_{\star} = 2.9 \times10^5 M_{\odot}$ \citep{Mcconnachie2012} and much shorter SFH of Draco, with 90\% of the stars being formed more than 10 Gyrs ago \citep{Aparicio2001}, with respect to Fornax, that has a stellar mass $M_{\star} = 2 \times10^7 M_{\odot}$ \citep{Mcconnachie2012} and is dominated by stars younger than 10 Gyrs \citep{deboer2012}.

As for Sculptor, most studies find that it formed most of its stars in the first 2-3 Gyrs of evolution and that it has been passive since then \citep{Gallart2015, Bettinelli2019, delosreyes2022}, but there are some works with slightly different results \citep{deboer2012}. This dSph has a stellar mass $M_{\star} = 2.3 \times10^6 M_{\odot}$ \citep{Mcconnachie2012}\footnote{We note that the stellar masses quoted  assume a mass-to-light ratio of unity in V-band (Vega magnitudes), independently on the galaxy stellar populations, and should therefore be considered as approximate.}, and a DM halo mass within 3 kpc of around $M_{halo} = 3 \times10^8 M_{\odot}$ \citep{Hayashi2020}. This would yield an upper limit to $M_{\star}/M_{halo}$ of $\approx 7.7 \times 10^{-3}$, in the regime where the initial DM cusp could be softened by stellar feedback. However, most works cannot differentiate between a core and a cusp \citep{Breddels2013a, Breddels2013b, Richardson2014, Zhu2016, Strigari2017, Hayashi2020} \citepalias[][hereafter]{Breddels2013a, Breddels2013b, Richardson2014, Zhu2016, Strigari2017, Hayashi2020}, some show preference for a cored DM halo \citep{Battaglia2008, Walker2011,Amorisco2012, Kaplinghat2019} \citepalias{Battaglia2008, Walker2011,Amorisco2012, Kaplinghat2019}, and others infer a cuspy DM halo \citep{Read2019, exposito-marquez2023} \citepalias{Read2019,exposito-marquez2023}. In general, obtaining more accurate measurements of the DM distribution in dSphs would impose stringent observational constraints on DM models, significantly advancing our understanding of the nature of DM. 

An advantageous characteristic of Sculptor is that it is known for having at least two distinct stellar components \citep{Tolstoy2004}, a metal-rich (MR) central component with velocity dispersion of around 6 \kms, and a more extended metal-poor (MP) population with velocity dispersion around 11 \kms \citep{Battaglia2008,Walker2011,ArroyoPolonio2024}. These two populations also occupy different locations in the energy versus angular momentum plane according to the Schwarzschild modeling \citep{Schwarzschild1979} performed by \cite{Breddels2013a}. Using these two different populations as independent tracers of the underlying potential has shown to be a very powerful tool to constrain the DM density profile \citep{Battaglia2008, Walker2011}, as it allows alleviating the known mass-velocity anisotropy degeneracy \citep{Binney1982}. 

The determination of DM density profiles in dSphs mostly relies on the dynamical modeling techniques applied to reproduce the structure and kinematics, provided that the galaxy is in dynamical equilibrium. In \cite{Iorio2019}, the authors run $N$-body simulations aimed at assessing the impact of tidal disturbances in Sculptor, reproducing carefully Sculptor's stellar structure, internal kinematic properties and using orbits based on the 3D motion derived by Gaia DR2 data. Even on orbits with repeated pericentric passages as internal as 35 kpc, they find that the stellar component was not significantly affected by the tidal field of the MW and that the system remained in dynamical equilibrium in the past 2 Gyr; hence the observed stellar kinematics serve as a robust tracer of the internal dynamics. This condition should in principle be fulfilled also on the orbits inferred by the more recent Gaia eDR3 data when accounting also for the perturbation induced by the infall of the Large Magellanic Cloud \citep{Battaglia2022a}, which suggest that Sculptor might have experienced its first pericenter passage around the Milky Way (MW) about 0.5 Gyrs ago at a distance of around 50 kpc. 

Under the assumption of dynamical equilibrium and describing the dSph as a combination of collisionless components, the system can be fully described in terms of distribution functions (DFs). In this work, we make use of DFs that depend on the action integrals to determine the internal structure of the DM density distribution of Sculptor. These DFs have been widely tested and used to describe the dynamics of a wide variety of systems as dSphs, globular clusters or the MW \citep[for example,][]{Binney2015,Pascale2018,Pascale2019,VAsiliev2019b,Read2021, Binney2023, dellacroce2024}. For our analysis, we use the largest sample of homogeneously derived line-of-sight (l.o.s.) velocities and metallicities ([Fe/H]) available for this galaxy \citep{Tolstoy2023}. We employ an unsupervised and self-tuning classification scheme based on a mixture modelling approach, in which stars are probabilistically assigned to different populations whose properties are determined by maximizing the likelihood of the entire observed dataset. Additionally, we include a component describing possible sources of contamination. By doing so, we test the robustness of the presence of the third component found by \citet[][hereafter AP24]{ArroyoPolonio2024}, potentially serving as a confirmation of this component.

The paper is organized as follows: In Sec.~\ref{sec:Data} we present the main characteristics of the dataset. In Sec.~\ref{sec:Methodology} we present the models and the methods used to explore the parameter space and to assign a probability of membership to each star. In Sec.~\ref{sec:Results_Sculptor_main} we show our inferences for the DM density profile of Sculptor and the properties of Sculptor' stellar components. In Sec.~\ref{sec:Mock} we test our methodology on a mock galaxy with a cuspy DM halo. In Sec.~\ref{sec:comparison_literature} we compare our results with the literature. In Sec.~\ref{sec:Discussion} we discuss the physical implications of our results. Finally, in Sec.~\ref{sec:conclusions} we summarize the main results and conclusions of our work.

%--------------------------------------------------------------------
\section{Data} \label{sec:Data}
In this work, we make use of the $\vlos$ and metallicity ([Fe/H]) measurements of individual stars presented in \cite{Tolstoy2023}. Here, we summarize the main characteristics of the dataset, for further details about the data reduction process we refer the reader to \cite{Tolstoy2023}. The dataset is a homogeneous collection of measurements acquired at the Very Large Telescope (VLT) using the GIRAFFE spectrograph in Medusa mode and the FLAMES instrument. The LR8 grating was used, covering wavelengths between 820.6 and 940.0 nm with a spectral resolving power of 6500 \citep{Pasquini2002}. This region was used as it includes the near-IR Ca II triplet (CaT). This feature is a well-known and tested indicator of the metallicity of red giant branch (RGB) stars, also in composite stellar populations \citep{Battaglia2008, Starkenburg2010, Carrera2013,Vasquez2015}. 

The data-set is composed of 67 independent observations of 44 pointings homogeneously reduced and analyzed. A zero-point calibration, based on a global shift applied to the spectra for each individual field, was made to ensure that there were no velocity offsets between different exposures.  In total, there are 1604 stars highly likely members to Sculptor with reliable $\vlos$, out of those, 1339 have also reliable [Fe/H]. In our analysis, we include only the subsample of stars with available metallicity measurements. The velocities have a mean error of $\pm$0.6 \kms,\footnote{It has been shown that the velocity uncertainties of the single exposures must be slightly calibrated \citepalias{ArroyoPolonio2024}. However, this calibration affects very little the averaged velocities, and it has no effect on the global parameters of the galaxy.} while the mean metallicity error is $\pm$0.1 dex. This is the largest and more extended homogeneous dataset used to compute the dark matter density profile of the Sculptor dwarf galaxy.

As we will discuss in Sec.~\ref{sec:Methodology}, the DF-based models used in this study assume spherical symmetry, whereas Sculptor appears flattened in the plane of the sky, with an ellipticity $e = 1 - b/a$ = 0.32 \citep{Muñoz2018}, where $b$ is the length of the projected minor axis and $a$, the one of the projected major axis. It is common to compare dSph data to spherical models \citep[][for example]{Walker2011, Read2019}. In this work, we investigate the impact of this assumption by exploring different mappings that translate the flattened stellar distribution of Sculptor into a spherical one. Let $\{x,y\}_i$ be the position on the plane of the sky of the $i$-th star, with the reference x-axis aligned with the galaxy major axis. We define
\begin{equation}\label{eq:smaxisradius}
    R_i = \sqrt{x_i^2+\left(\frac{y_i}{1-e}\right)^2},
\end{equation}
which we refer to as the "semi-major axis radius". This would be our reference parameter to compare with spherical models. In order to characterize the scatter in the results produced by this choice, we will also repeat the analysis assuming a "circular radius": 
\begin{equation}\label{eq:circular}
    R_{i,c} = \sqrt{x_i^2+y_i^2},
\end{equation}
and a "circularized" radius, i.e. the radius of the circle with the same area as the ellipse defined by the position of the $i-$th star and the ellipticity of Sculptor, this is: 
\begin{equation}\label{eq:circularized}
    R_{i,cz} = \sqrt{x_i^2(1-e)+\left(\frac{y_i^2}{1-e}\right)}.
\end{equation}

In this article, we adopt the results based on the semi-major axis radius obtained using Eq.~(\ref{eq:smaxisradius}) as a reference. However, as we explore in Sec.~\ref{sec: different_radii} how the inference on the DM density distribution of Sculptor depends on this choice. We adopt the values of the center, ellipticity and position angle by \cite{Muñoz2018} and distance modulus by \cite{MartinezVazquez2015}.

The large angular size of Sculptor causes the 3D systemic velocity of the galaxy to project differently over the lines-of-sight to different stars \citep{Walker2008}. We remove this perspective $\vlos$ gradient using the formulae presented in the appendix of \cite{Walker2008}. To do so, we adopt the systemic proper motion of Sculptor ($\mu_{\alpha , *}$, $\mu_{\delta}$) = (0.1, $-$0.147) [$mas$ $yr^{-1}$], which are already corrected for the zero-point offsets measured with quasars located within 7 degrees from Sculptor \citep{Battaglia2022a}. We adopt a systemic l.o.s. velocity of 111.2 \kms \citepalias{ArroyoPolonio2024}, which we keep fixed throughout the analysis.

Even though Sculptor is mostly pressure supported, some statistically significant l.o.s. velocity gradients have been detected in the past and ascribed to internal rotation \citep{Battaglia2008, Zhu2016}. However, the most recent works, which can now fully account for the artificial velocity gradient induced by the 3D relative motion of Sculptor and the Sun calculated on accurate and precise Gaia DR3 proper motions, and use more recent data-sets, find either inconclusive or no signs of rotation \citep{Martinez-garcia2023,ArroyoPolonio2024}.

\section{Methodology} 
\label{sec:Methodology}

In this section, we present the chemo-dynamical models used, and the methodology applied to compare them with the observed data. 

In Sculptor, in past dynamical modeling works, the stellar mass have been found to be at least one order of magnitude lower than the dynamical one at all radii \citep[see][and references therein]{Battaglia2022b}. Therefore, we will assume that the gravitational potential is generated by the DM component only. The stellar component is modelled as a superposition of two distinct stellar populations that are tracers of the gravitational potential and that are represented by action-based DFs. We refer to these as the MR and MP populations. We use a model with multiple components as in \citetalias{ArroyoPolonio2024} the authors have shown that Sculptor is better described by such a model rather than by a single-population one with a metallicity gradient. In addition, we add a third stellar component aimed at modeling possible "contamination". This choice is motivated by the fact that \citetalias{ArroyoPolonio2024}, using the same dataset, found that Sculptor is best described by a model that includes, in addition to the dominant MR and MP populations, a third, much smaller component. This component, detected also in \cite{Tolstoy2023}, consists of approximately 20 stars, exhibiting a systemic l.o.s. velocity shift of 15 \kms relative to the velocity of Sculptor. Such a systemic velocity shift is hard to explain as a component in equilibrium. Thus, we account for the possible presence of these stars in the model as a third population of stars that does not trace the total potential. We refer to these as pop-3.

In Sec.~\ref{sec:met_Models}, we present the action-based models used to describe the stellar phase-space distribution and the model used for describing the DM component. In Sec.~\ref{sec:met_Bayesian_inference}, we introduce the adopted posterior distribution, describe the priors, and provide details on the Bayesian inference used to estimate the models' free parameters, as well as the methodology employed to compare the models with the data. Finally, in Sec.~\ref{sec:met_Pmembership}, we show how we assign the probability of membership of each star to each population. In App.~\ref{app:met_MCMC}, we summarize the free parameters in our models and the specifics of the MCMC runs used to explore the parameter space.

\subsection{Models} \label{sec:met_Models}

In Sec.~\ref{subsec:met_dm}, we describe the models we use for the DM component. In Sec.~\ref{subsec:met_sd}, we present the DF-based models for the MR and MP stellar populations of Sculptor, as well as the non-DF based distribution used for the third "contamination" component. Finally, in Sec.~\ref{sec:met_Bayesian_inference} we outline the Bayesian method used for parameter estimation.

\subsubsection{DM component} \label{subsec:met_dm}

DM is described as a fixed component whose potential is generated by the density distribution: 
\begin{equation} \label{eq:DM}
   \begin{aligned}
& \rho(r)=\rho_0\left(\frac{r}{r_s}\right)^{-\gamma}\left[1+\left(\frac{r}{r_s}\right)^\alpha\right]^{\frac{\gamma-\eta}{\alpha}} \exp \left[-\left(\frac{r}{r_{\text {cut }}}\right)^{\xi}\right],
\end{aligned}
\end{equation}
The above density distribution is that of a double power-law model with an exponential truncation in the outer parts. $\rs$ is a scale radius and it indicates the location of the transition between the inner part ($r < \rs$), where the power-law index is $\gamma$, and the outer part ($\rs < r < \rcut$), where the power-law index is $\eta$; $\alpha$ controls the steepness of the transition between those slopes; $\rho_0$ is a characteristic density; $\rcut$ and $\xi$ define the position and the strength of an exponential cut-off. We fix these two parameters to $\rcut = 20$ kpc and $\xi = 1$. The only effect of the truncation is to ensure that the DM has finite mass when $\eta\le3$, $\rcut = 20$ kpc is chosen because it is approximately ten times greater than the distance of our outermost star, and therefore it does not have a significant impact on the inference. With a proper normalization, we substitute $\rho_0$ by the total DM mass $\MDM$ \footnote{The total DM mass can be computed as $\MDM =\int_{r=0}^{\infty}\rho(r)r^2dr$.}. Note that, due to extrapolation and the arbitrary cut-off, this parameter by itself is not informative. In summary, the DM component is described by 5 free parameters \{$\MDM,\rs,\alpha,\eta,\gamma$\}.

\subsubsection{Stellar components}\label{subsec:met_sd}

The phase-space distribution of the MP and MR populations of Sculptor is described by the action-based DFs \citep{Vasiliev2019}
\begin{equation} \label{eq:df}
    \begin{aligned}
    \fpop(\JJ) & =\frac{\foi \Mi}{(2 \pi \Joi)^3}\left[1+\left(\frac{\Joi}{\hi(\JJ)}\right)\right]^{\Gammai}\left[1+\left(\frac{\gi(\JJ)}{\Joi}\right)\right]^{-\Bi}, \\
    \gi(\JJ) & \equiv \gri J_r+\gzi J_z+\gphii \left|J_\phi\right|, \\
    \hi(\JJ) & \equiv \hri J_r+\hzi J_z+\hphii\left|J_\phi\right|.
    \end{aligned}
\end{equation}
The DF in Eq.~(\ref{eq:df}) is a generalisation of the family of $f(\JJ)$ DFs introduced by \cite{Posti2015}. When $p={\rm MP}$, Eq.~(\ref{eq:df}) refers to the MP population DF and its parameters, while when $p={\rm MR}$ it refers to the MR population DF and its associated parameters. In the DFs~(\ref{eq:df}), $J_r$, $J_z$ and $J_\phi$ are the actions associated to the radial, vertical and azimuthal directions; $\foi$ is a factor that normalizes each DF to the corresponding total mass $\Mi$; $\Joi$ is a characteristic action scale; $\Gammai$ and $\Bi$ are dimensionless parameters setting the asymptotic behavior of the DF for $|\JJ|$ larger and smaller than $\Joi$, respectively. When $|\JJ|\gg \Joi$ the DF of the $p$-th is a power-law of index $-\Bi$, when $|\JJ|\ll \Joi$ is a power-law of index $-\Gammai$. These parameters produce similar effects on the density distribution for one-component systems; finally, the parameters $\hri$, $\hzi$, $\hphii$, and $\gri$, $\gzi$, $\gphii$ in the $\hi(\JJ)$ and $\gi(\JJ)$ functions, mostly control the velocity anisotropy in the inner and outer parts of the model, respectively. We impose $\hri+\hzi+\hphii=3$ and $\gri+\gzi+\gphii=3$ \citep{Binney2014}. In addition, due to spherical symmetry, we impose $\gzi=\gphii$ and $\hzi=\hphii$, so that the DFs (\ref{eq:df}) depend only on the radial action $J_r$ and the angular momentum $L\equiv J_z+|J_\phi|$. These choices leave us with only two free parameters, which we choose to be $\hzi$ and $\gzi$. These conditions could be easily relaxed to allow axisymmetry in our models. However, we consider that axisymmetric models have not been sufficiently tested to be robustly applied to observed data, as opposed to spherically symmetric modeling. We plan to perform axisymmetric modeling in a future contribution, once the methodology has been more extensively tested in this context, which is a work carried out in another study (Gherghinescu et al. in prep).

We consider stars as mere tracers of the total potential, so we do not account for them when solving the Poisson equation. As a consequence, the masses $\Mi$ of the MR and MP components are not free parameters. Therefore, to describe the two main stellar components, we have 10 free parameters, namely \{$\JMP,\GammaMP,\BMP,\gzMP,\hzMP, \JMR,\GammaMR,\BMR,\gzMR,\hzMR$\}.

The angle-action formalism, the implementation of the MR and MP DFs, and the computation of all relevant DF integrals are performed using the Action-based Galaxy Modeling Architecture (AGAMA) software library \citep{Vasiliev2018, Vasiliev2019}. AGAMA is designed for the dynamical modeling of stellar systems and provides efficient algorithms for computing gravitational potentials, DFs, and orbit integrations. The library is particularly well-suited for action-based dynamical modeling, offering methods for computing actions and angles in gravitational potentials, as well as tools for constructing self-consistent equilibrium models of galaxies and star clusters.

We introduce $\LL^p$ as the probability distribution function describing the likelihood of a star to belong to the $p$-th population. This probability function encapsulates the complete structural and chemo-dynamical information of each population. Specifically, when $p=\MP$ or $\MR$
\begin{equation}\label{eq:MPMR}
    \LL^p \equiv \frac{\PP^p_d(R,\vlos)\,\PP^p_{\MM}(\met)\,\omega(R,G)}{2\pi\int\PP^p_d(R,\vlos)\,\omega(R,G) R \,\dd R\dd\vlos\dd G},
\end{equation}
where $G$ is the apparent $G$-band magnitude, $\PP^p_d(R,\vlos)$ is a marginalization of Eq.~(\ref{eq:df}) over the l.o.s. $z$ and ${\bf v}_\perp$, i.e. the component of the velocity vector perpendicular to the plane of the sky. We refer to $\PP^p_d(R,\vlos)$ as projected DF, and it represents the probability of finding a star on the plane of the sky at a the position $R$ from the galaxy center and with l.o.s. $\vlos$. More explicitly
\begin{equation}
    \PP_d^p(R,\vlos) \equiv \frac{\int\int \fpop(\JJ)\,\dd z \dd^2 {\bf v}_{\perp}}{\Mi}.
\end{equation}
We compute it by using a marginalisation by Monte Carlo integration approach implemented in AGAMA (see Gherghinescu et al. in prep for details). In Eq.~(\ref{eq:MPMR}), $\PP_{\MM}^p(\met)$ is the metallicity distribution function of the $p$-th population, which we assume to follow a Gaussian distribution
\begin{equation}\label{eq:Gaussian}
    \PP^p_\MM(\met)  =  \frac{1}{\sqrt{2\pi}\sigma_{\MM,p}}\exp{\left( -\frac{(\mathcal{M}_p - \text{[Fe/H]})^2}{2\sigma^2_{\MM,p}}\right)},
\end{equation}
while $\omega(R, G)$ is the selection function, for which we adopt the one from Eq.~2 of \citetalias{ArroyoPolonio2024}. In that work, the authors derived a continuous selection function using a Gaussian kernel smoothing approach, incorporating the same spectroscopic data used in this work, as well as a complete Gaia sample of confirmed astrometric members from \cite{Tolstoy2023}, identified using only Gaia DR3 data. We note that in Eq.~(\ref{eq:MPMR}), since the metallicity distribution $\PP_{\MM}(\met)$ does not depend on $R$ nor on $\vlos$ and it is a normalized Gaussian, we have neglected it in the denominator. Also, the denominator can be rearranged as $\int\int \PP^p_d(R, \vlos)\,\Omega(R) R \dd R \dd\vlos$, where $\Omega(R) = \int\omega(R,G)\dd G$. The metallicity distribution of the $\MP$ and $\MR$ populations add four more free parameters, namely \{$\MM_{\MP},\sigma_{\MM,\MP}, \MM_{\MR},\sigma_{\MM,\MR}$\}.

When $p$ labels the pop-3 population ($p=C$), 
\begin{equation}\label{for:3pop}
    \LL^\C = \frac{ \,\PP_{\VV}^\C(\vlos)\,\PP_{\MM}^\C(\met)\omega(R,G)}{2\pi\int \Omega(R)R\dd R},
\end{equation}
where $\PP^\C_{\MM}$ and $\PP^\C_{\VV}$ are Gaussians describing the velocity and metallicity distributions of the third population. The metallicity distribution function is the same as in equation~\ref{eq:Gaussian}, but with $p=\C$. For the l.o.s. velocity distribution of the third population we assume a Gaussian 
\begin{equation}\label{eq:vgauss}
    \PP^\C_\VV(\vlos)  =  \frac{1}{\sqrt{2\pi}\sigma_{\VV,\C}}\exp{\left( -\frac{(\VV_\C - \vlos)^2}{2\sigma^2_{\VV,\C}}\right)},
\end{equation}
adding four parameters more, the mean metallicity of the third population $\MM_\C$, and the standard deviation, $\sigma_{\MM,\C}$, the mean velocity of the velocity distribution $\VV_\C$, together with its standard deviation, $\sigma_{\VV,\C}$.

The denominators in Eqs.~(\ref{eq:MPMR}) and (\ref{for:3pop}) ensure that the DF-based MR and MP populations, as well as the spatial distribution of the third component, are normalized to have a unit integral over the selected region. This region is defined as a circle with a radius of 2 kpc, corresponding to the distance of the farthest star from the galaxy center, centered on Sculptor nominal center.

\subsection{Bayesian inference}
\label{sec:met_Bayesian_inference}

We fit the data within a full Bayesian framework. Let $\boldsymbol{\theta}$ be the vector of free parameters, and define the set of observables for the $i$-th star as $\bzeta_i = \{R_i, v_{los,i}, \Delta v_{los,i}, \met_i, \Delta\met_i\}$, where $i = 1, \ldots, \nstar$, with $\nstar=1339$ (see Sec.~\ref{sec:Data}), we define the likelihood
\begin{equation} \label{eq:likelihood}
    \LL(\boldsymbol{\theta}) = \prod_i^{\nstar}\left(\sum_pf_p\,(\LL^p\ast\EE)(\bzeta_i)\right )
    ,
\end{equation}
where $\ast$ indicates the convolution with the error function $\EE$, which in our case is the product of two Gaussians with null mean and dispersions equal to $\Delta v_{los,i}$ and $\Delta\met_i$. The metallicity distribution function is always Gaussian, so the resulting convolution is Gaussian as well. The velocity distribution is Gaussian only for the third population, whereas for the MP and MR populations, the integral is evaluated numerically and performed directly by \textsc{AGAMA}. In Eq.~(\ref{eq:likelihood}) the index $p$ runs over all the populations ($p=\MP,\MR,\C$) and the index $i$ over all the stars in our sample; $f_p$ is the fraction of stars that belongs to population $p$, where \{$\fMP$, $\fMR$\} are free parameters, while $f_\C = 1-\fMP-\fMR$. In total we have 25 free parameters, specifically
\begin{equation}\begin{aligned}
    \btheta \ \equiv \ & \{\MDM, \rs, \alpha, \eta, \gamma, \\ & \JMP,\GammaMP,\BMP,\gzMP,\hzMP,\MM_{\MP},\sigma_{\MM,\MP}, \\
    & \JMR,\GammaMR,\BMR,\gzMR,\hzMR,\MM_{\MR},\sigma_{\MM,\MR}, \\
    &
    \VV_\C,\sigma_{\VV,\C}\MM_\C,\sigma_{\MM,\C}, \\
    & \fMP, \fMR\}
\end{aligned}\end{equation}

%   priors
We adopt uniform priors on all the parameters explored in this work. Specifically, we sample the parameters 
$\JMP$, $\JMR$ and $\MDM$, in logarithmic space, as they are expected to span several orders of magnitude. The remaining parameters are sampled in linear space, where the values are expected to vary within a more constrained range. The posterior distribution follows from Bayes' theorem and is given by the product of the uniform priors and the likelihood function (equation~\ref{eq:likelihood}). Further details on the MCMC fitting procedure are given in Appendix~\ref{app:met_MCMC}, while the results of the fit and the priors used are summarized in Table~\ref{tab:fitThreepop}.

\subsection{Membership probabilities} \label{sec:met_Pmembership}
Making use of the formalism presented above, we can assign a probability of each star to belong to a certain population. Following \citetalias{ArroyoPolonio2024}, we define the probability of the $i-$th star to belong to the population $p$ as 

\begin{equation} \label{eq:pmem}
    P_i^p = \frac{f_p (\LL^p\ast\EE)(\bzeta_i)}{\sum_k f_k (\LL^k\ast\EE)(\bzeta_i)}.
\end{equation}
In the above equation $k$ runs over all the stellar components, so $k=\MR, \MP, 3$. In our Bayesian approach, we determine the probability of membership as follows: we select 5000 models based on the posterior probability distribution (PPD) and calculate the membership probability for each model. Then, we take the median of these probabilities and scale it so that the total probability for each star across all populations equals 1. This scaled median serves as the final indicator of the membership probability.

\section{Results}\label{sec:Results_Sculptor_main}

After exploring the parameter space, we have constrained all 25 free parameters of the models, characterizing the three populations and the DM density distribution. The PPDs for the parameters of the stellar components and the DM, independently, are shown in App.~\ref{app:cornerpolots}. The median values along with confidence intervals in are also listed in Tab.~\ref{tab:fitThreepop}. From now on, when a parameter is quoted with errors the values indicate the median and the 1-$\sigma$ confidence interval. For reference, we also show some important parameters of these models in Tab.~\ref{tab:extra-parameters}. In Sect.~\ref{sec:stellar_populations}, we present the main properties of the stellar components and compare them with the literature; in Sec.~\ref{sec:dark_matter}, we analyze the DM density profile inferred for Sculptor; finally, in Sec.~\ref{sec: different_radii}, we analyze the sensitivity of the method to different choices for the observational radius used to compare with the data. In App.~\ref{sec:individualpop}, we show the results of applying the methodology to each of the individual main populations independently.

\begin{table*}[h]
\centering
\caption{Inference on models free parameters for the fit to Sculptor dSph data. Details on the procedure are given in Sec.~\ref{sec:Methodology}.}
\renewcommand{\arraystretch}{1.25}
\begin{tabular}{c|lccccc}
\hline
\hline
Component & Parameter & Prior & Median & $1\sigma$ & $3\sigma$ \\
\hline
\hline
& $\log \JMP$ [kpc\,\kms]  & [0.3,2.5]          & 1.63 & [1.27, 1.85] & [0.88, 2.21] \\
& $\hzMP$                  & [0, 1.5]           & 0.56 & [0.31, 0.84] & [0.03, 1.41] \\
& $\gzMP$                  & [0, 1.5]           & 1.07 & [0.95, 1.17] & [0.79, 1.34] \\
MP & $\GammaMP$            & [0, 3]             & 0.81 & [0.49, 1.04] & [0.01, 1.42] \\
& $\BMP$                   & [3, 30]            & 13.10 & [8.09, 22.51] & [5.77, 29.79]\\
& $\MM_\MP$ [dex]          & [-2.7, -2.1]       & -2.000 & [-2.016, -1.976] & [-2.067, -1.943] \\
& $\sigma_{\MM,\MP}$ [dex] & [0.15, 0.45]       & 0.309 & [0.297, 0.320] & [0.279, 0.344] \\
& $\fMP$                   & [0.5, 0.75]        & 0.66 & [0.63, 0.69] & [0.56, 0.74] \\

\hline
& $\log\JMR$ [kpc\,\kms]   & [0.3,2.5]             & 1.21 & [0.98, 1.34] & [0.71, 1.49] \\
& $\hzMR$                  & [0, 1.5]              & 0.99 & [0.51, 1.35] & [0.01, 1.49] \\
& $\gzMR$                  & [0, 1.5]              & 0.98 & [0.89, 1.05] & [0.75, 1.19] \\
MR & $\GammaMR$            & [0, 3]                & 0.24 & [0.07, 0.49] & [0.00, 0.95] \\
& $\BMR$                   & [3, 30]               & 21.6 & [13.9, 27.8] & [8.9, 29.9]\\
& $\MM_{\MR}$ [dex]        & [-1.65, -1.3]         & -1.415  & [-1.443, -1.390] & [-1.500, -1.344] \\
& $\sigma_{\MM,\MR}$ [dex] & [0.15, 0.45]          & 0.253  & [0.237, 0.270] & [0.211, 0.308] \\
& $\fMR$                        & [0.24, 0.5]      & 0.32 & [0.29, 0.35] & [0.24, 0.41] \\

\hline
& $\VV_\C$ [\kms]          & [-30, 30]     & 13.6   & [11.6, 15.7] & [5.6, 20.2] \\
& $\sigma_{\VV,\C}$ [\kms] & [0, 20]       & 6.9    & [5.5, 9.2] & [3.8, 15.7] \\
pop-3 & $\MM_\C$ [dex]     & [-4.5, -1.3]  & -2.81  & [-3.04, -2.56] & [-3.39, -2.16] \\
& $\sigma_{\MM,\C}$ [dex]  & [0.15, 0.90]  &  0.53  & [0.39, 0.67] & [0.20, 0.90] \\
& $f_\C$                    & -            & 0.016  & [0.011, 0.022] & [0.006, 0.042] \\

\hline
& $\log \MDM$ [$M_{\odot}$] & [7, 11]  & 9.0    & [8.7, 9.4] & [8.4, 10.0] \\
& $\log \rs$ [kpc]          & [-3, 1]  & -0.10  & [-0.21, 0.07] & [-0.35, 0.50] \\
DM & $\alpha$               & [0, 7]   & 3.7    & [2.1, 5.8] & [0.9, 6.9] \\
& $\eta$                    & [2, 7]   & 3.2    & [2.4, 3.9] & [2.0, 5.0] \\
& $\gamma$                  & [0, 1.9] & 0.39   & [0.13, 0.62] & [0.00, 0.95] \\
\hline

\hline
\hline
\end{tabular}
\tablefoot{The table is divided in four parts. The first three show the parameters of the MP, MR and the third stellar populations, while the forth part shows the inference on the parameters of the DM component. Col.~1 indicates the components the parameters are referred to; Col.~2 lists the parameters; Col.~3 shows the range of uniform priors we used in the EMCEE runs; Cols.~4,~5 and~6 lists the median values, $1\sigma$ and $3\sigma$ confidence intervals. Details on the MCMC procedure and on how the 1- and 3-$\sigma$ confidence intervals are computed are given in App.~\ref{app:met_MCMC}.}\label{tab:fitThreepop}
\end{table*}

\begin{table}[h]
\centering
\caption{Inference on other interesting parameters for the fit to Sculptor dSph data.}
\renewcommand{\arraystretch}{1.25}
\begin{tabular}{c|cccc}
\hline
\hline
Parameter & Median & $1\sigma$ & $3\sigma$ \\
\hline
$\gamma_{0.5r_h}$ & 0.43 & [0.19, 0.64] & [0.01, 0.98] \\
$\gamma_{r_h}$ & 0.53 & [0.33, 0.72] & [0.07, 1.00] \\
$\gamma_{2r_h}$ & 1.27 & [0.92, 1.68] & [0.43, 2.69] \\
$\log(\rho_{150})$ & 8.01 & [7.95, 8.06] & [7.83, 8.16] \\
$\log([M_\odot/\text{kpc}^{-3}])$\\
$\log(M_{r_h} [M_{\odot}])$ & 7.01 & [6.97, 7.05] & [6.87, 7.13] \\
$\log(M_{\rm 2kpc} [M_{\odot}])$ & 8.70 & [8.54, 8.85] & [8.33, 9.17] \\
$(M_{halo}/M_{\star})_{r_h}$ & 13.3 & [12.1, 15.6] & [9.9, 17.4] \\
$(M_{halo}/M_{\star})_{2\text{kpc}}$ & 154 & [122, 200] & [84, 351] \\
$J(0.5^{\circ})$ & 18.15 & [18.03, 18.26] & [17.84, 18.54] \\
$[\text{Gev}^2/\text{cm}^{-5}]$ \\
$D(0.5^{\circ})$ & 18.07 & [17.97, 18.17] & [17.82, 18.39] \\
$[\text{Gev}/\text{cm}^{-2}]$ \\

\hline
\end{tabular}
\tablefoot{Median and confidence inteval values for the following parameters: $\gamma_{xr_h}$ is the inner slope at $x$ times the 3D half-light radius \citep[$0.32^{+0.02}_{-0.02}$ kpc][]{Muñoz2018}; $\rho_{150}$ is the density at 150 pc; $M_{r_h}$ is the mass enclosed within at the 3D half-light radius; $M_{\rm 2\text{kpc}}$ is the cumulative DM mass at 2kpc, the position at which we have our outermost star; $(M_{halo}/M_{\star})_{2kpc}$ and $(M_{halo}/M_{\star})_{r_h}$ are the ratio between DM halo mass and stellar mass at the 3D half-light radius and at 2 kpc; finally, $J(0.5^{\circ})$ and $D(0.5^{\circ})$ are the annihilation and decay factors (see Sec.~\ref{sec:Dis_JD}).} \label{tab:extra-parameters}

\end{table}

\subsection{Properties of the stellar populations} \label{sec:stellar_populations}

In Sec.~\ref{sec:stellar_populations_MP/MR} we describe the properties of the main two, DF-based, stellar components. In Sec.~\ref{sec:stellar_populations_thirdcomponent} we describe the properties of the third, non DF-based, stellar component.

\subsubsection{MR and MP stellar components} \label{sec:stellar_populations_MP/MR}

According to our model results, the MR population has a mean metallicity of $\MM_{\MR} = -1.42^{+0.02}_{-0.03}$ dex and metallicity dispersion of $\sigma_{\MM,\MR} = 0.253^{+0.017}_{-0.016}$ dex, while the MP component has $\MM_{\MP} = -2.000_{-0.016}^{+0.024}$ dex and $\sigma_{\MM,\MP} = 0.309^{+0.011}_{-0.014}$ dex. These values are in agreement, within 1-$\sigma$, with those provided by \cite{Tolstoy2004} and \citetalias{ArroyoPolonio2024}. 

\begin{figure*}
    \centering
        \includegraphics[width=2\columnwidth]{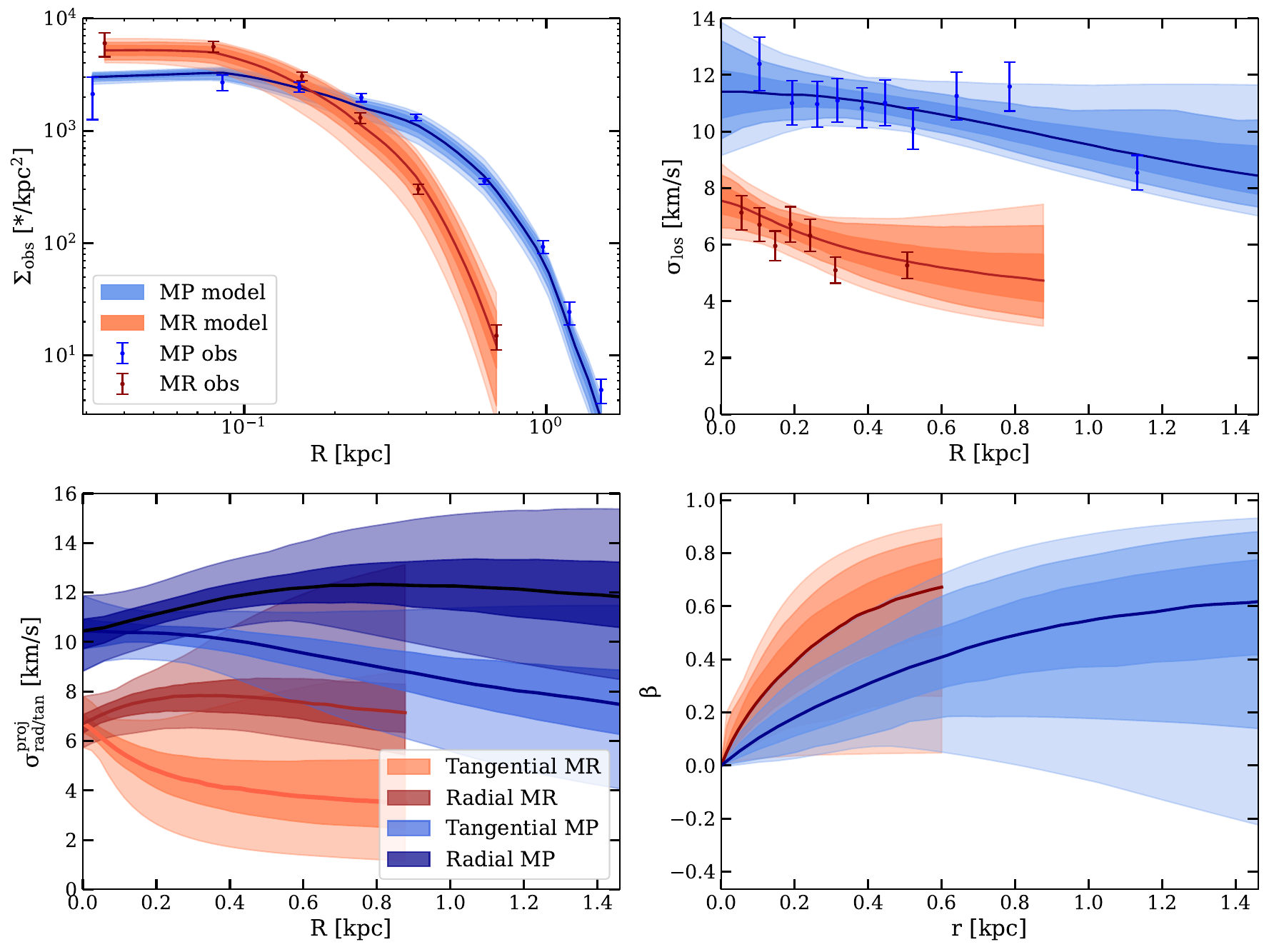}
        \caption{Comparison between the median models of the MP and MR populations of Sculptor and the observed binned data from \citetalias{ArroyoPolonio2024}. Points indicate the observed data with the bars indicating 1-$\sigma$ uncertainties, the solid lines the median models and the colored bands the 1-, 2- and 3-$\sigma$ ranges. The MP and MR components are shown in blue and red, respectively, as indicated in the legend. Upper left panel: observed surface density profiles. Upper right panel: l.o.s. velocity dispersion profiles. Lower left panel: radial and tangential projected velocity dispersion profiles, only 
        1- and 3-$\sigma$ bands are shown for simplicity. Lower right panel: 3D velocity anisotropy profiles.} \label{Fig:Observed_prifles}% Total $\chi**2$ = 40
        
    \end{figure*}

The upper panels of Fig.~\ref{Fig:Observed_prifles} shows the observed projected density\footnote{Observed projected density means that for the models we show $\Omega(R)\Sigma_{int}(R)$, where $\Sigma_{int}$ is the intrinsic-projected density, so that we fairly compare with the data.} (top left) and the l.o.s. velocity dispersion profiles (top right) computed from the dataset, alongside the median, 1-, 2- and 3-$\sigma$ bands on the corresponding quantities produced from the models, with the aim of evaluating how accurately the DFs reproduce the observed properties of the MR and MP components. We emphasize that we do not fit directly these binned profiles, rather we fit individual star velocity and metallicity. The profiles are only shown as a comparison. To compute the binned l.o.s. velocity dispersion profiles, we have assumed probability membership from \citetalias{ArroyoPolonio2024}, accounting for the uncertainties in $v_{los}$. From the top left panel of Fig.~\ref{Fig:Observed_prifles}, we note that the density profile of the MP population decreases in the inner region. This decline, however, is due to the selection function (Eq.~\ref{eq:likelihood}) rather than an intrinsic drop in density within the galaxy's central parts. Even if it can not be directly seen in this figure, if we compare the surface density profiles produced by our models to the Plummer profiles \citep{Plummer1911} obtained by \citetalias{ArroyoPolonio2024}, we find that the MR component is more concentrated and does not follow a Plummer profile, while the MP component is more extended and similar to a Plummer profile.

The bottom right panel of Fig.~\ref{Fig:Observed_prifles} shows the velocity anisotropy parameter profile of the MP and MR populations. It is defined as $\beta = 1 - \frac{\sigma^2_t}{\sigma^2_r}$, where $\sigma^2_r$ and $\sigma^2_t$ indicate the radial and the tangential velocity dispersions in 3D, respectively. Our DF-based models allow us to compute the velocity anisotropy profiles by integrating the second order velocity moments, as a natural output of the model. We find that both populations are isotropic in the central regions, followed by a transition to radial anisotropy. This transition occurs more rapidly in the MR component, whereas in the MP population, the increase in radial anisotropy is more gradual, reaching a lower median value of $\beta$, but with large uncertainties. This is broadly consistent with the results of \citetalias{Battaglia2008},  \citetalias{Amorisco2012} and \citetalias{Strigari2017}, but it is in tension with the results of \citetalias{Zhu2016} who find that the MR component is isotropic (on average, $\beta \simeq 0$) and the MP component tangential (on average, $\beta \simeq -0.5$). If we focus on works that model Sculptor with one population only, \citetalias{Hayashi2020} find a velocity distribution radially biased at all radii  (on average, $\beta \simeq 0.4$), as also \citetalias{Read2019} (on average, $\beta \simeq 0.3$), \citetalias{Pascalethesis} isotropic velocity distribution (on average, $\beta \sim 0$) and \citetalias{Breddels2013a} tangentially biased velocity distribution (on average, $\beta \simeq -1$). By analyzing the three-dimensional kinematics of a limited number of stars in Sculptor, \cite{Massari2018} find a preference for radial orbits in Sculptor beyond the core radius ($\beta = 0.86^{+0.12}_{-0.83}$), and \cite{delpino2022} find also radial orbits with large uncertainties ($\beta = 0.46\pm0.44$) along the major axis of Sculptor. In general, despite significant uncertainties, the literature shows no clear consensus on velocity anisotropy, although most studies tend to favor isotropic or radial orbits over tangential ones. The comparison of $\beta$ with the literature is expanded in Sec.~\ref{sec:comparison_literature}.

In the lower left panel of Fig.~\ref{Fig:Observed_prifles}, we present predictions for the radial and tangential velocity dispersions projected in the plane of the sky, defined as $\sigma^{\text{proj}}_{\text{rad}}$ and $\sigma^{\text{proj}}_{\text{tan}}$, respectively. The figure shows that $\sigma^{\text{proj}}_{\text{tan}}$ decreases while $\sigma^{\text{proj}}_{\text{rad}}$ increases as $R$ increases for both populations; this can serve as a prediction for future proper motion studies.

\begin{figure*}
    \centering
        \includegraphics[width=1.85\columnwidth]{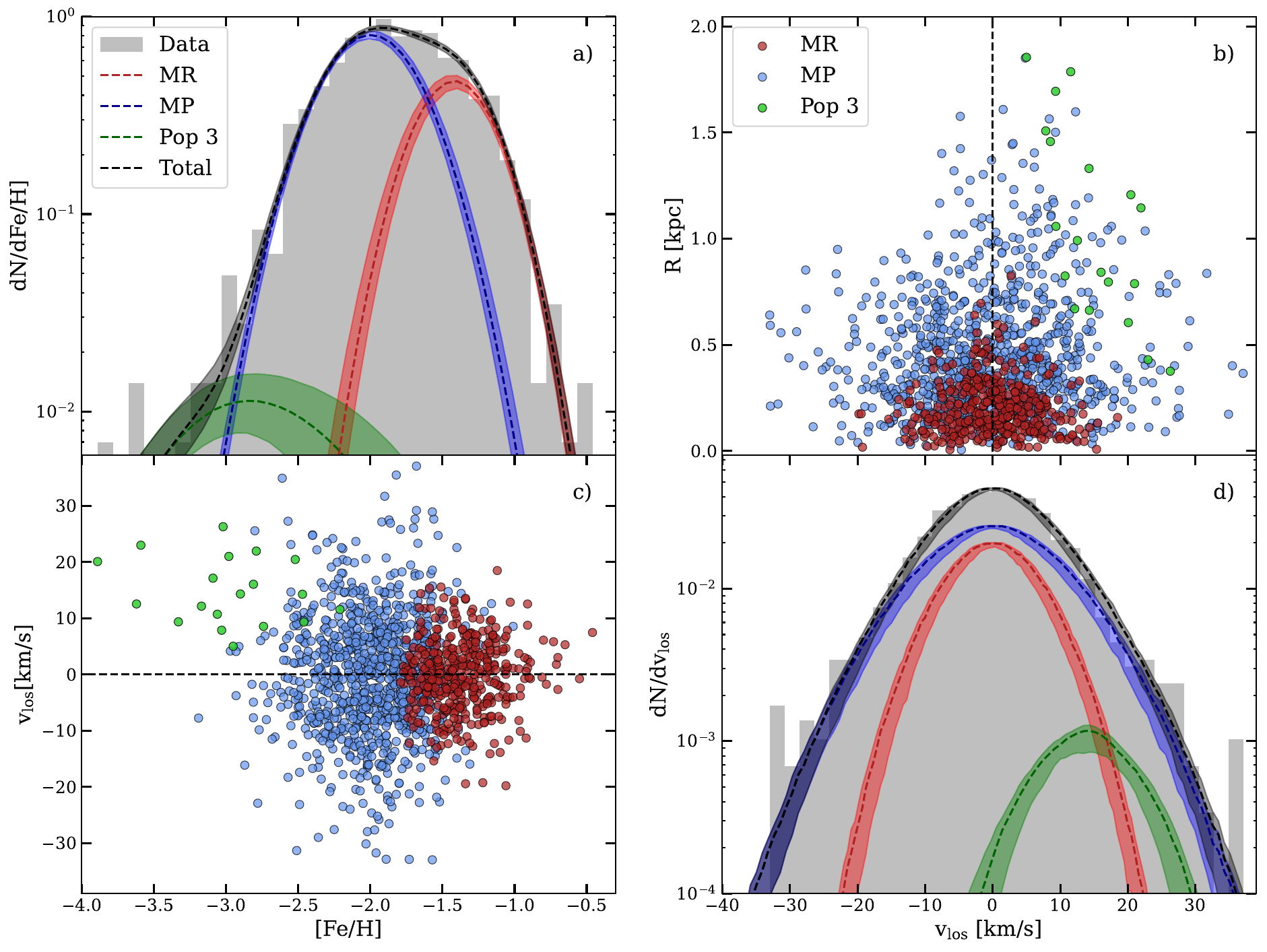}
        \caption{ Chemo-dynamical stellar populations of Sculptor. a) / d) panels: Normalized metallicity/l.o.s. velocity distribution of the spectroscopically observed member stars of Sculptor alongside the median models (dashed colored lines) resulting from the fit of Sec.\ref{sec:stellar_populations}. Red blue and green indicate the MR, MP components and third populations, respectively. Black shows the sum of the three components, as indicated in the legend. Bands indicate the 1-$\sigma$ confidence intervals. b) / c) panels: Distribution of stars in the $v_{los}$ vs [Fe/H] / $R$ vs $v_{los}$ planes. In each panel, the stars are color-coded by the population to which they have the highest probability of belonging, red for the MR, blue for the MP and green for the third population, as indicated in the legend. The systemic velocity of Sculptor is indicated with straight dashed lines.}
        \label{Fig:members}
        
    \end{figure*}

As we explained in Sec.~\ref{sec:met_Pmembership}, we can assign each star a membership probability for each population. In Fig.~\ref{Fig:members} we show the distribution of the stars in the $\vlos$ vs $\met$ and $R$ vs $\vlos$ planes, and we leverage these probabilities to color-code stars according to the population to which they have the highest probability of belonging (see equation~\ref{eq:pmem}), alongside with the overall $\vlos$ and [Fe/H] distributions of our models. A comparison between Fig.~\ref{Fig:members} and Fig. 3 from \citetalias{ArroyoPolonio2024} reveals a generally consistent identification of the populations. Nevertheless, some minor differences are present. The model we find to describe the MR population (bottom right panel) has a higher overall l.o.s. velocity dispersion, 7.0$^{+0.2}_{-0.2}$ \kms with respect to the results of \citetalias{ArroyoPolonio2024}, 6.3$^{+0.3}_{-0.3}$ \kms and it is more concentrated in the center. This can be seen also on the upper-right panel of Fig.~\ref{Fig:Observed_prifles} in which the models for the MR component show slighlty larger $\slos$ than the binned data.  %\gbcom{how much higher and at what radius?}{\bf agree. Say numbers. Also, we already said that the MR pop is more concentrated, and you say it again for the third time in the next sentence... Also (personally) I find annoying comparing figures between papers.}. 
Therefore, we find that even though the simple distributions used in \cite{Walker2011, Pace2020, ArroyoPolonio2024} work well to distinguish between populations, they might introduce biases in the membership of the stars that could affect the dynamical modeling if the membership is not re-evaluated as we do in this work.

\subsubsection{Third stellar component} \label{sec:stellar_populations_thirdcomponent}

We identify a third stellar component, consistent with the one reported by \citetalias{ArroyoPolonio2024} and seen also in \cite{Tolstoy2023}. Specifically, the mean metallicity of the metallicity distribution function is $\MM_\C = -2.81^{+0.25}_{-0.23}$ dex, with a dispersion $\sigma_{\MM,\C} = 0.53^{+0.14}_{-0.14}$ dex. These values are in very good agreement with the ones from \citetalias{ArroyoPolonio2024} ($\MM_\C = -2.9^{+0.3}_{-0.2}$ dex and $\sigma_{\MM,\C} = 0.5^{+0.10}_{-0.10}$ dex, respectively). Regarding the velocity distribution, we find a mean velocity offset with respect to Sculptor of $\VV_\C=13.6^{+2.1}_{-2.0}$ \kms and a velocity dispersion of $\sigma_{\VV,\C} = 6.9^{+2.3}_{-1.4}$ \kms, again consistent with those of \citetalias{ArroyoPolonio2024} ($\VV_\C= 14.3^{+2.4}_{-2.6}$ \kms and $\sigma_{\VV,\C} = 7.7^{+2.3}_{-1.7}$ \kms). 
 
The median fraction of stars expected to belong to this component is $f_\C=0.016$, which corresponds to a number of stars $f_\C \nstar=21$, with a 3-$\sigma$ range $f_\C=[0.006, 0.042]$ (corresponding to [8, 56] stars), in very good agreement with the values from \citetalias{ArroyoPolonio2024}, a median of $f_\C=0.017$ and 3-$\sigma$ range $f_\C=[0.007, 0.058]$. According to the stars with larger probability of belonging to the third component than to the MP or MR ones (Eq.~\ref{eq:pmem}), we find 19 stars out of the 24 initially found by \citetalias{ArroyoPolonio2024}. Moreover, we have to consider that these DFs produce physically motivated models.

Modeling the metallicity distribution function (MDF) of a galaxy is challenging due to the complex chemical enrichment processes involved. As a result, Gaussian distributions may not adequately represent the MDF, potentially leading to asymmetric distributions, as commonly observed in dwarf galaxies \citep[e.g.][]{Kirby2011,Leaman2012,Taibi2022}. However, the case of Scuptor is somewhat peculiar. Indeed, the stars of the 3rd population that exhibit low metallicity, representing the metal-poor tail of the MP population, also display a peculiar kinematic signature, with a notably large velocity offset. This combination of low metallicity and peculiar kinematics suggests that it may represent a genuinely distinct third population. We refer the reader to \citetalias{ArroyoPolonio2024} for a discussion on the possible origin of this component.

\subsection{Dark matter density profile}\label{sec:dark_matter}

\begin{figure} 
    \centering
        \includegraphics[width=0.9\columnwidth]{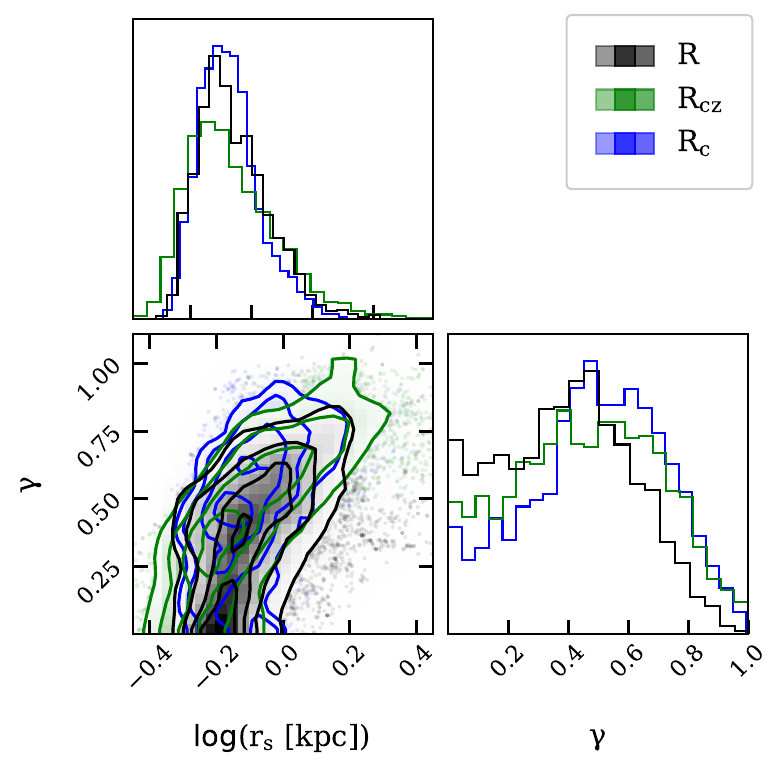}
        \caption{Two dimensional marginalized posterior distribution of $\rs$ and $\gamma$ and the corresponding one-dimensional posterior distributions. The PPDs are shown  when using semi-major axis radius (black), circularized radius (green) and circular radius (blue), as indicated in the legend.}
        \label{fig:gamma_dif_rad}
    \end{figure}

Fig.~\ref{fig:gamma_dif_rad} presents the marginalized two-dimensional and one-dimensional posterior distributions for the parameters $\gamma$ and $\rs$. The full marginalized posterior distributions for the remaining DM distribution parameters are shown in Fig.~\ref{Fig:cornerplot_DM}. We first focus on the results obtained using the semi-major axis radius (represented by the black lines in Fig.~\ref{fig:gamma_dif_rad}), with a detailed discussion of other cases provided in Sec.~\ref{sec: different_radii}. The PPD for $\gamma$ is slightly skewed towards $\gamma = 0$, suggesting a somewhat constant DM density (core) at the galaxy's center, even though considering the entire PPD Sculptor does not have a clear constant-density region. We can also notice that $\rs$ is degenerate with $\gamma$; the larger $\rs$, the steeper $\gamma$. The median values and 1-$\sigma$ confidence errors for the logarithmic inner slope and the scale radius are $\gamma = 0.39^{+0.23}_{-0.26}$ and $\rs=0.79^{+0.38}_{-0.17}$ kpc, more than twice the 3D half-light radius of the stellar component, $0.32^{+0.02}_{-0.02}$ kpc \citep{Muñoz2018}. 

\begin{figure}
    \centering
        \includegraphics[width=1\columnwidth]{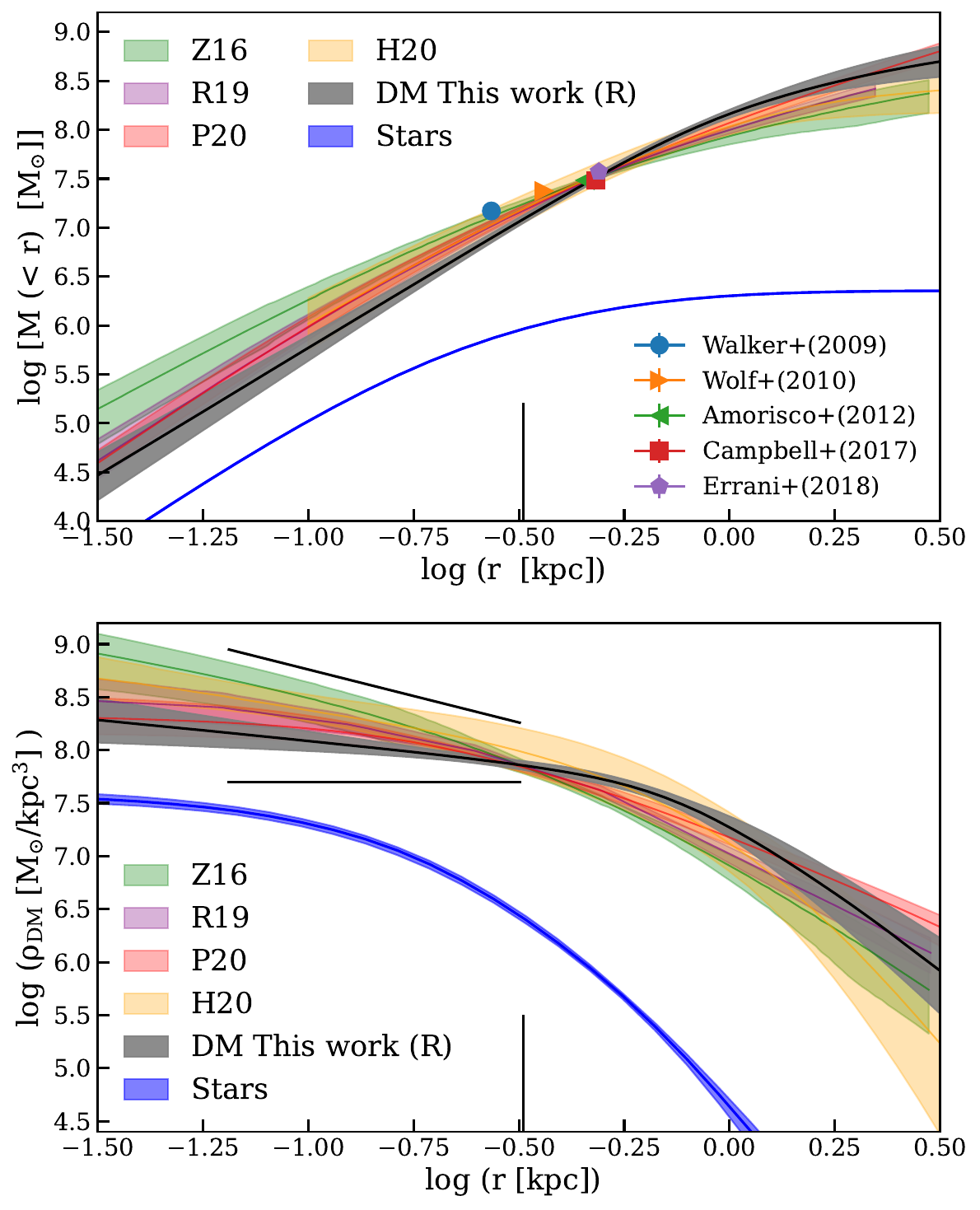}
        \caption{DM halo profiles inferred for Sculptor. Upper panel: Median enclosed halo mass profile together with $1\sigma$ confidence intervals (black bands). Mass indicators are shown with error bars, color-coded according to the legend. Lower panel: DM halo density profile. For reference, the two inner lines indicate slopes of 1 (cusp) and 0 (core). In our dataset, there are 15 stars with $\log(R\,[{\rm \text{kpc}}])<-1.5$, i.e. below the lowest radius plotted. The stellar mass profile is shown with a blue bands. The works highlighted in the legends are \citetalias{Zhu2016}, \citetalias{Read2019}, \citetalias{Pascalethesis} and \citetalias{Hayashi2020}.}
        \label{fig:Cumulat_den_profiles}
    \end{figure}

In Fig.~\ref{fig:Cumulat_den_profiles} we show the median cumulative DM halo mass and density profiles together with the 1- and 3-$\sigma$ confidence intervals from our models, and we compare them to the ones in the literature. Our profiles are in good agreement with the ones from the literature at almost all radii. Nevertheless, as already anticipated, in the density profile we find a deficit of DM with respect to a NFW profile in the inner parts, which is more pronounced compared with other works. This could be partially because of the lower uncertainties presented in our work, the new high-quality dataset and the inclusion of the discrete 2 population modeling with DFs. In Sec.~\ref{sec:comparison_literature} we will compare our results with the literature ones in more detail. We also show the distributions for the stellar component, assumed to be the sum of the MR and the MP populations. To compute it, we assume the total stellar mass to be $2.3 \times 10^6 $M$_\odot$ \citep{Mcconnachie2012}. It is important to note that this value is computed assuming a mass-to-light ratio of 1 $M_{\odot}/L_{\odot}$ in V-band \footnote{However, this value could be different for Sculptor. For example, a synthetic population based on Basti \citep{Basti2013} stellar evolutionary models and a Salpeter initial mass function for a constant SFH betwen 12-13 Gyr ago and a MDF centered in -2.3 with a spread of 0.5 dex result into a mass-to-light ratio around 1.68 in V-band. But a different choice of the initial mass function could lead to a distinct value.}. The density of the stellar component decreases more rapidly than that of the DM, resulting in a DM-to-stellar mass ratio ($M_{halo}/M_{\star}$) of approximately 6 at the center within 30 pc. This ratio increases to around 13 at the 3D half-light radius and reaches 154 at a distance of 2 kpc, where the farthest star in our dataset is located. Therefore, this system is highly DM dominated. However, its central regions are less DM dominated compared to previous studies. 

As mentioned in the introduction, the inference of constant density cores in the DM density profile of some galaxies has motivated alternative DM models. For example, in SIDM, DM particles thermalize due to collisions in the central regions of halos, leading in some cases to the formation of constant-density cores \citep{Sameie2020}. In fuzzy DM, the DM particles are so light that they behave as waves, reaching equilibrium in a soliton configuration and forming cores \citep{Tulin2018}. These profiles are more consistent with our findings than a pure NFW profile. However, based on our inference of the semi-major axis radius and the likelihood ratio test statistics, we can also reject a pure core ($\gamma = 0$) larger than 0.7 kpc at the 99\% confidence level, which is a useful constraint for these DM theories \citep[see ][]{Chen2017}.

\subsection{Impact of different choices of radius} \label{sec: different_radii}

So far, we have compared the spherically symmetric models to the observed data by accounting for the flattened shape of Sculptor's stellar component by labeling stars with the semi-major axis radius of the ellipse defined by the position of that star and the ellipticity of the galaxy (eq.~\ref{eq:smaxisradius}). However, different works use different definitions (eqs.~\ref{eq:circular} and \ref{eq:circularized}). Therefore, we have repeated the whole fitting procedure considering them. 

Fig.~\ref{fig:gamma_dif_rad} shows how the inference of Sculptor' DM halo inner logarithm slope and scale radius are affected by these choices. The PPDs are fully consistent with each other, overlapping in the parameter space. This is reflected in the median values of $\gamma$ and $\rs$, which are $\gamma = 0.50^{+0.25}_{-0.30}$ and $\rs = 0.78^{+0.26}_{-0.17}$ kpc when using the circular radius, and $\gamma = 0.45^{+0.20}_{-0.35}$ and $\rs = 0.77^{+0.40}_{-0.23}$ kpc  for the circularized radius. Both results are in very good agreement with the inference for the semi-major axis radius, which is $\gamma = 0.39^{+0.23}_{-0.26}$ and $\rs=0.79^{+0.38}_{-0.17}$ kpc, but yielding a slightly steeper inner logarithmic slope. This result suggests that the assumption of sphericity does not significantly impact our findings, as accounting for the system's ellipticity in different approaches—or neglecting it—yields consistent results. However, this does not eliminate the need for comparisons with axisymmetric models, which can more accurately capture the effects of Sculptor’s flattening. In the literature, there have been attempts to apply axisymmetric models to real data, with methods based on Jeans modeling \citep{Zhu2016,Hayashi2020}; however, Jeans methods do not guarantee to provide physical solutions, as some of them could be generated by negative DFs, and have their own caveats. Perhaps the biggest difference between the assumptions is related to the $\log(\rho_{150})$ value quoted in Tab.~\ref{tab:extra-parameters}. For the circular radius, it is $\log(\rho_{150} [M_\odot/\text{kpc}^{-3}]) = 8.14^{+0.04}_{-0.04}$, while for the circularized radius, it is $\log(\rho_{150} [M_\odot/\text{kpc}^{-3}]) = 8.07^{+0.05}_{-0.05}$. It is also important to note that the semi-major axis radius is systematically larger than the circular and circularized ones. Therefore the value quoted in this case is lower than in the other cases. 

\begin{figure}
        \includegraphics[width=1\columnwidth]{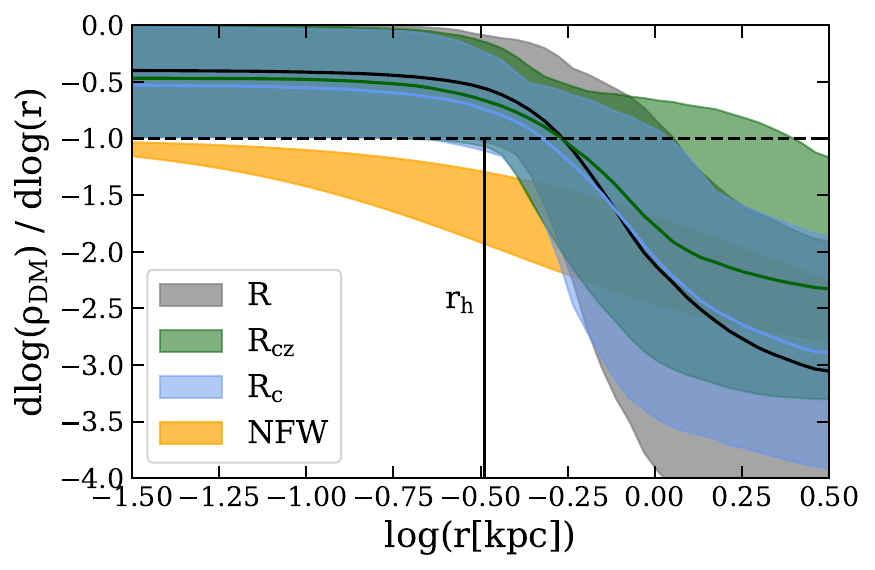}
        \caption{Log-slope of the DM density profile for Sculptor. The gray, green and blue bands show the inferred DM density according to our models for the $R$, $R_{cz}$ and $R_{c}$ cases for a 3-$\sigma$ confidence level. The orange band indicates the log slope profile of a pure NFW with mass and concentration as expected from cosmological consideration (see text for details), as indicated in the legend. The vertical line shows the 3-D half-light radius as reference.}
        \label{fig:logrhoprofile}
    \end{figure}

An interesting result that we obtain is that we can exclude a pure cusp in all the cases, and more generally a NFW profile, at a 3-$\sigma$ confidence level over a significant range of radii. This can be seen in Fig.~\ref{fig:logrhoprofile}, where we compare the logarithmic slope profile obtained in the three cases with the expectations for a NFW model assuming halo masses between $10^8M_{\odot}$ to $10^{10}M_{\odot}$ consistent with the scatter in the stellar-to-halo mass relation \citep{Sales2022} and the relation between the concentration of the NFW profile and its total mass for satellites from \cite{Moline2023} at redshift $z=0$. The inconsistency with respect to a NFW profile arises not only from the value of $\gamma$ but also from the steepness of the transition $\alpha$. Our model predicts a median value for $\alpha$ of 3.6 and a 1-$\sigma$ confidence interval of [2.1, 5.8], whereas for a NFW  $\alpha =$ 1, meaning the transition between the two slopes ocurrs very rapidly. So, although we cannot confirm the existence of a core in this galaxy, we can robustly exclude a NFW DM density profile. This can possibly indicate an impact of baryonic effects in modifying the DM halo density profile of this galaxy. It is important to note that our prior on $\gamma$ only allows physical models with $\gamma \geq 0$, which could affect our inference as the prior is centered on cuspy profiles while the cored models lie on the borders \citep[see Sec. 4.6 of][]{Read2018}. We expect that adopting a prior on $\gamma$ more balanced between cored and cuspy profiles would only strengthen our conclusions.

\section{Test on a Sculptor-like mock} \label{sec:Mock}

We apply our methodology to a mock data-set of $R$, $\vlos$ and [Fe/H] for a spherically symmetric 2-component stellar system resembling Sculptor, embedding it in a cuspy spherical DM halo. The same test for a cored halo is presented in App.~\ref{app:mock}. This experiment is meant to show that, with the quality of the dataset available, we would be able to find a cusp in Sculptor if it were there. However, as we use the same DFs to generate and analyze the mock, this test would not reveal systematic biases due to our choice of the DF. A similar experiment for a more comprehensive sets of mocks, was conducted by \cite{Read2021} using different DFs, but in the single-stellar component case.

We generate the mock data-set from the DF parameters listed in Tab.~\ref{tab:Mock-test}, which have been chosen so that the populations have different velocity dispersion profiles that resemble those of Sculptor. To assign metallicities to each stellar population, we use the MDFs of the metal rich and metal-poor components of Sculptor reported in \citetalias{ArroyoPolonio2024}. Following this, we extract the velocities, positions and metallicities for a set of 1339 stars. We extract them accounting for the same errors they have in our observational dataset, we do this by adding Gaussian noise to the quantities. Only the two stellar components used for the dynamical modeling are considered in this analysis, as we aim to test whether the information we have for these two components is sufficient to detect a cusp in Sculptor using our methodology.

\begin{figure}
        \includegraphics[width=1\columnwidth]{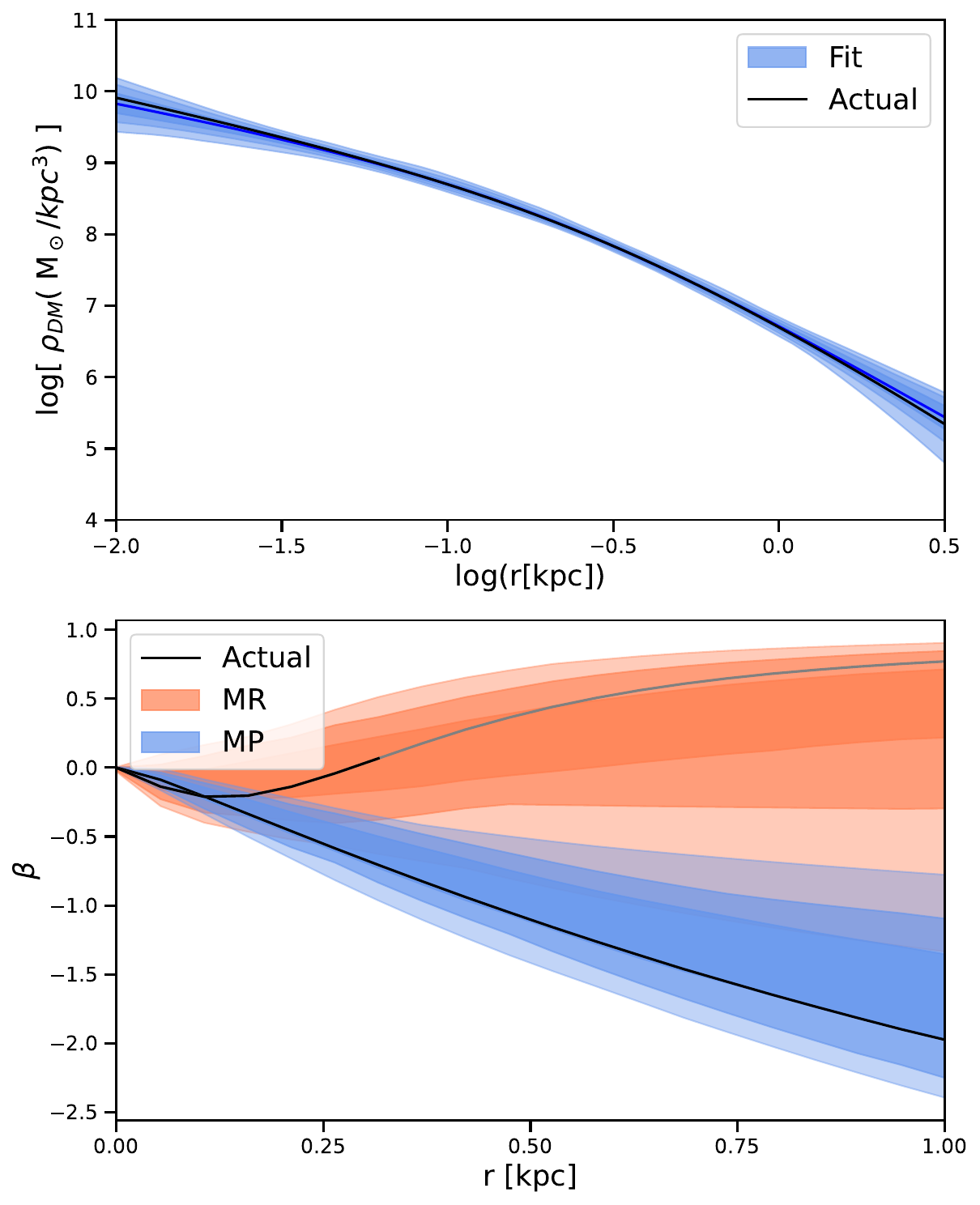}
        \caption{Upper panel: DM density profile of the mock galaxy. In black, we show the underlying one, the median of the predicted profile is shown with a solid blue line and the blue bands indicate the 1-, 2- and 3-$\sigma$ confidence intervals. Lower panel: Velocity anisotropy profile for the MP and MR stellar components. As solid lines, we show the values for the underlying model and in bands the best-fit for the MR and the MP component, indicating the 1-, 2- and 3-$\sigma$ confidence intervals. The solid black line corresponding to the MR component become gray in the region in which we do not produce stars that belong to that component.}
        \label{fig:Mock_2pop}
    \end{figure}

The median and the 3-$\sigma$ range of the PPD for the inferred parameters after running our dynamical modeling are listed in Tab.~\ref{tab:Mock-test}. We can see that the inferred parameters are in very good agreement with the true values, which are very close to the median and always within 3-$\sigma$. The DM density and anisotropy profiles of the mock data-set and the ones resulting from the fitting procedure are shown in Fig.~\ref{fig:Mock_2pop}. We recover with high precision the true profiles, within the 1-$\sigma$ bands. Therefore, we conclude that, at least in this ideal case where we build our mock with the same DF we are using to analyze it, our model is capable of recovering the cusp in a dSph-like galaxy with a dataset similar to the one of Sculptor. Furthermore, it can independently constrain the anisotropy profile of both populations. However, the uncertainties on the inner logarithm slope do not allow to discard a cored DM distribution within 3-$\sigma$.

When analyzing a cored halo mock (see App.~\ref{app:mock}) the inferred parameters also show very good agreement, the true values are always very close to the medians and lie within 3-$\sigma$. Even though the inference of $\gamma$ is consistent with 0 only within 3-$\sigma$, it is something to expect as we only allow models with $\gamma > 0$, but the PPD is completely skewed towards 0. This same effect could explain while $r_s$ is also in the 3-$\sigma$ border, as it is degenerated with $\gamma$. Overall, the inferred profiles also match the true ones with high precision.

\begin{table*}[h]
\centering
\caption{Inference on models free parameters for the cuspy mock.}
\renewcommand{\arraystretch}{1.25}
\begin{tabular}{c|lcccccc}
\hline
\hline
Component & Parameter & Prior & Median & $1\sigma$ & $3\sigma$ & Actual\\
\hline
\hline
& $\log \JMP$   & [0.3,2.5]                       & 0.92 & [0.69, 1.09] & [0.44, 1.25] & 0.8\\
& $\hzMP$      & [0, 1.5]                        & 0.27 & [0.15, 0.46] & [0.05, 0.99] & 0.05\\
& $\gzMP$      & [0, 1.5]                        & 1.05 & [0.93, 1.16] & [0.55, 1.33] & 1.2\\
MP & $\GammaMP$   & [0, 3]                          & 1.15 & [0.93, 1.31] & [0.65, 1.58] & 1 \\
& $\BMP$ & [3, 25]                       & 17.0 & [11.9, 22.3] & [7.5, 24.9] & 14\\
& $\MM_{\MP}$ [dex]  & [-2.1, -1.7]          & $-$2.00 & [-2.02, -1.99] & [-2.04, -1.97] & -2 \\
& $\sigma_{\MM,\MP}$ [dex] & [0.15, 0.45]  & 0.300 & [-0.29, -0.31] & [0.27, 0.33] & 0.3\\
& $\fMP$                        & [0.5, 0.75]        & 0.58 & [0.56, 0.60] & [0.54, 0.64] & 0.6\\

\hline
& $\log \JMR$   & [0.3,2.5]                       & 1.47 & [1.29, 1.61] & [1.04, 1.81] & 1.5\\
& $\hzMR$      & [0, 1.5]                        & 0.47 & [0.15, 1.04] & [0.01, 1.48] & 0.6\\
& $\gzMR$      & [0, 1.5]                        & 0.47 & [0.40, 0.54] & [0.32, 0.78] & 0.4\\
MR & $\GammaMR$   & [0, 3]                          & 0.33 & [0.13, 0.61] & [0.01, 0.97]& 0.4\\
& $\BMR$ & [3, 25]                       & 17.4 & [13.5, 22.1] & [10.0, 24.9] & 20\\
& $\MM_{\MR}$ [dex]  & [-1.65, -1.3]         & $-$1.43  & [-1.44, -1.41] & [-1.47, -1.39] & -1.4\\
& $\sigma_{\MM,\MR}$ [dex] & [0.15, 0.45]  & 0.30  & [0.29, 0.31] & [0.27, 0.33] & 0.3\\
& $\fMR$                        & [0.25, 0.5]        & 0.42 & [0.40, 0.44] & [0.36, 0.46] & 0.4\\

\hline
& $\log \MDM$ [$M_{\odot}$]  & [7, 11]  & 8.6    & [8.4, 8.8] & [8.1, 9.1] & 8.5\\
& $\log \rs$ [kpc]             & [-3, 1]  & -0.53  & [-0.75, -0.26] & [-1.43, 0.03] & -0.5\\
DM & $\alpha$                      & [0, 7]   & 1.8    & [0.8, 4.2] & [0.4, 6.9] & 1\\
& $\eta$                       & [2, 7]   & 2.65    & [2.38, 3.39] & [2.07, 4.79] & 3\\
& $\gamma$                      & [0, 1.9] & 1.18   & [0.67, 1.40] & [0.02, 1.57] & 1\\
\hline
& $\log \MDM$ [$M_{\odot}$]  & [7, 11]  & 9.03    & [8.95, 9.09] & [8.70, 9.15] & 8.5\\
& $\log \rs$ [kpc]             & [-3, 1]  & -1.23  & [-1.33, -1.13] & [-1.54, -0.93] & -0.5\\
DM 1-pop & $\alpha$              & [0, 7]   & 3.4    & [2.0, 5.5] & [1.1, 7.0] & 1\\
& $\eta$                       & [2, 7]   & 2.06    & [2.01, 2.14] & [2.00, 2.40] & 3\\
& $\gamma$                      & [0, 1.9] & 0.37   & [0.11, 0.70] & [0.00, 1.04] & 1\\
\hline
\hline
\hline
\end{tabular}
\tablefoot{The table is divided in four parts. The first three show the parameters of the MP, MR stellar populations and the DM component of the two populations analysis. The last part shows the parameters for the DM component of the single population analysis. Col.~1 indicates the components the parameters are referred to; Col.~2 lists the parameters; Col.~3 shows the range of uniform priors we used in the EMCEE runs; Cols.~4,~5 and~6 lists the median values, $1\sigma$ and $3\sigma$ confidence intervals; and finally Col.~7 shows the actual value used to produce the mock. Details on the MCMC procedure and on how the 1- and 3-$\sigma$ confidence intervals are computed are given in App.~\ref{app:met_MCMC}}\label{tab:Mock-test}
\end{table*}

We also analyzed the cusped mock galaxy assuming that there is only one single population, resembling the observational scenario where the two independent populations are not considered. The results of the fit are listed in Tab.~\ref{tab:Mock-test}. The inferred parameters are offset from the true ones and not within 1-$\sigma$. Notably, now we infer a distribution much closer to a core, but with large uncertainties and consistent with a cusp within 3-$\sigma$.

\section{Comparison with the literature} \label{sec:comparison_literature}

Here, we compare our results with those from the literature. Specifically, in Sec.~\ref{sec:compa_masses}, we compare the mass estimators used to compute the total mass within a given radius. In Sec.~\ref{sec:comparision_dyn}, we present the DM density profiles reported in previous works and compare them with ours.

\subsection{Mass estimators}\label{sec:compa_masses}

In the literature, many authors have managed to constrain the dynamical mass enclosed within a certain characteristic radius by using mass estimators of the form: 
\begin{equation}\label{eq:massest}
    M_{dyn}(< xR_h) = \frac{K \Rh\langle\sigma_ {l.o.s.}\rangle^2}{G}.
\end{equation}
In the above equation, $M_{dyn}(< x\Rh)$ is the dynamical mass within $x$ times the projected half-light radius $\Rh$ (other formulations express it as a function of the 3-D half-light radius), $K$ is a dimensionless constant, $G$ the gravitational constant, and $\langle\sigma_ {l.o.s.}\rangle$ the mean l.o.s. velocity dispersion of the stellar system. Different studies have derived different values for $K$ and $x$, depending on their different assumptions. In the cumulative mass profile shown in the top panel of Fig.~\ref{fig:Cumulat_den_profiles} one can compare our results with the ones obtained by using the estimators derived by \cite{Walker2009a,Wolf2010, Amorisco2012, Campbell2017, Errani2018}. The estimators that provide a value of the enclosed mass beyond the projected half-light radius are in very good agreement with our results \citep{Amorisco2012, Campbell2017, Errani2018}. On the other hand, estimators that provide a value of enclosed mass around the half-light radius \citep{Walker2009a, Wolf2010} deviate more from our findings.

In general, simple mass estimators also show lower errors than the dynamical modelling methods when computed via error propagation. However, this might be a problem of calibration, as different works testing mass estimators in simulations have shown that there is larger scatter in the results \citep[][Sarrato et al. in prep]{gonzalez-samaniego2017} than uncertainties computed here by applying error propagation of the $M_{dyn}(< n\Rh)$ and $\Rh$ uncertainties.

\subsection{Dynamical modeling}\label{sec:comparision_dyn}

There are different techniques to estimate the DM density profile of a pressure-supported system \citep[see e.g.][]{Battaglia2022b}. Those that directly solve the Jeans equation are known as Jeans modeling \citep[][for example]{Read2019, Hayashi2020, Wardana2024}. Another method is Schwarzschild orbit-superposition modeling \citep{Schwarzschild1979}, which is based on representing the target galaxy as the superposition of orbits \citep{Breddels2013a, Breddels2013b}. Each orbit type is assigned a weight that is optimized to match the observed data. DF modeling techniques, such as the one used here, model the target phase-space distribution directly assuming an analytic form of the DF. This latter method has the advantage that all observables can be directly computed from the DF. These three methods have been applied to Sculptor in several works, and here we conduct a comprehensive comparison with all existing studies that have inferred the DM density of Sculptor. Tab.~\ref{tab:comparison} lists all the works and their main characteristics and assumptions. In general, our results are consistent with the existing literature, even though the quoted uncertainties in the different methods are large. Overall, most of the studies favor a core over a cusp for Sculptor. Only \citetalias{Read2019} and \citetalias{exposito-marquez2023} clearly favor cuspy solutions. Given the many differences between studies in terms of datasets, methods, and assumptions, it is challenging to identify the factors driving the discrepancies in the results. Nevertheless, in the following, we briefly describe each of them and compare their results with ours.

\renewcommand{\arraystretch}{1.2} % Stretch rows by 1.5 times the default height
\begin{table*}[] 
\centering
\caption{Literature results for the logarithm inner slope of the DM halo of Sculptor.}
\begin{tabular}{|l|lllllll|} 

          \hline 
           & 2-pop & Method     & Dataset     & Anisotropy   & Data      & $\gamma$ flex &  result  \\ \hline
\citetalias{Battaglia2008}       & YES    & Jeans      & B08                & OM rad(r)       & Binned    & NO            & Both fit, core preferred \\
\citetalias{Walker2011}       & YES   & Estimator  & W09                & constant     & Discrete  & -             & $\gamma =$0.05$^{-0.39}_{-0.51}$ \\
\citetalias{Amorisco2012}      & YES    & DF  & B08               & rad(r)       & Binned    & NO            & Both fit, core preferred \\
\citetalias{Breddels2013a}      & NO    & Schwar.    & W09 \& B08          & Free         & Binned  & YES           & Unconstrained    \\
\citetalias{Breddels2013b}      & NO    & Schwar.    &  W09 \& B08                & Free         & Binned  & NO            & Unconstrained    \\
\citetalias{Richardson2014}      & NO    & Schwar.    & W09           & Free         & Discrete  & YES            & Unconstrained    \\
\citetalias{Zhu2016}       & YES   & Axi. Jeans & W09                & Constant     & Discrete  & YES           & $\gamma =$0.5$\pm$0.3    \\
\citetalias{Strigari2017}       & YES   & DF  & W09 / B08        & Tang/Rad (r) & Binned  & NO            & Depends on data-set  \\
\citetalias{Kaplinghat2019}       & NO    & Jeans      & W09                  & Free         & Binned    & NO            & Both fit core preferred  \\
\citetalias{Read2019}       & NO    & Jeans      & W09                  & Free         & Binned    & YES           & $\gamma =$0.83$^{+0.30}_{-0.25}$      \\
\citetalias{Hayashi2020}       & NO    & Axi. Jeans & W09                  & Constant     & Binned    & YES           & $\gamma =$0.45$^{+0.41}_{-0.31}$      \\
\citetalias{Pascalethesis}       & NO    & DF  & I19                  & Free         & Binned    & NO            & Core       \\
\citetalias{exposito-marquez2023}      & NO    & M.L.       & W09                  & -            & Discrete  & -             & Cusp         \\
This work & YES   & DF  & T23               & Free         & Discrete  & YES           & $\gamma =$0.39$^{+0.23}_{-0.26}$     \\ 
\hline

\end{tabular}
\tablefoot{Col.~1 lists the literature works. Col.~2 indicates whether the authors use a 2-population model or assume that all the stars belong a single population. Col.~3 lists the mass modeling method used, M.L. stands for machine learning. Col.~4 shows the dataset the authors used, \& indicates both datasets combined and / each one independently. Col.~5 lists the assumpsion made on the anisotropy profile, OM stands for Osipkov-Merritt profile \citep{Osipkov1979,Merrit1985} and rad and tang for only radial or only tangential orbits. Col.~6 indicates whether the analysis is done on the bases of binned data or discrete data. Col.~7 indicates whether the value assume for $\gamma$ in the potential is flexible or whether the authors only tried different cases and compare them. Finally, in Col.~8 we lists the result of that work.} \label{tab:comparison}
\end{table*}

\citetalias{Battaglia2008}: In this study, the authors performed Jeans modeling fitting both populations simultaneously using binned data and semi-major axis radius to label the stars. The authors assume that the stellar populations are tracers that follow a Plummer density profile for the MP component and a Sersic profile for the MR. For the DM component, the authors test two different models, a NFW and a pseudo-isothermal sphere, and two different velocity anisotropy profiles, an Osipkov-Merritt (OM) \citep{Osipkov1979,Merrit1985} and constant profiles, finding that the OM one was preferred to the others. They compare cuspy and cored models based on $\chi ^2$ values, finding that the cored ones reproduce the data better than the cusped. In particular, the NFW model predicts larger velocity dispersion for the MR component than the observed one. Their best-fit model has a core of 0.5 kpc, in marginal agreement with the scale radius we find, $\rs=0.79^{+0.38}_{-0.17}$ kpc. Note that the scale radius of our profile is degenerate with the value of $\gamma$, see Fig.~\ref{Fig:cornerplot_DM}. It is remarkable that, despite the differences in the methodology and dataset, the results are highly consistent. 

\citetalias{Walker2011}: In this work, the authors use an estimator for the inner slope of the DM profile, which is a function of the ratios between the velocity dispersion and the half-light radii of two stellar populations. They show that this estimator is biased towards predicting cuspy profiles by applying it to a mock sample. Their analysis assumes a Gaussian velocity distribution, with the spatial distribution modeled by a Plummer profile. Data from \cite{Walker2009a} is used, and incompleteness is taken into account. They excluded NFW or steeper profiles for Sculptor with a significance level of 99.8\% when using circular radii and of 93.9\% when using semi-major axis radii. Interestingly, they found more evidence of a core when using circular radii, contrary to our results. However, both results are in agreement with our inference, and this could be just an effect of the different datasets used.

\citetalias{Amorisco2012}: In this work, the authors used energy and angular momentum based DFs to model the data from \cite{Battaglia2008}, using semi-major axis radii. In particular, they used Michie-king DFs \citep{king1962,Michie1963}, assuming a positive-defined velocity anisotropy that can vary with the radius. They fit both populations simultaneously, treating them as independent tracers of the underlying potential, with the assumption that the stars are just dynamical tracers. For the DM component, they compared only two different models, a NFW and a cored NFW profile. The model selection is based on $\chi ^2$ values. They found that a cored distribution fit the data better compared to a cusped configuration. The scale radius for the cored model is 0.36 kpc, slightly smaller than ours, but completely consistent given the errors and the degeneracy with $\gamma$.

\citetalias{Breddels2013a} $\&$ \citetalias{Breddels2013b}: In these works, the authors performed Schwarzschild modeling to fit the second and fourth velocity moments. They combined the data from both \cite{Battaglia2008} and \cite{Walker2009a}, and they fit binned data using circular radius (Eq.~\ref{eq:circular}) to label the stars. They assume that all the stars belong to a single population and are tracers of the underlying potential. In \citetalias{Breddels2013a} the inner slope of their DM potential is a free parameter, while in \citetalias{Breddels2013b} the authors test different DM density profiles and compared them according to $\chi ^2$ values. In Schwarzchild models, the velocity anisotropy is an output, so it is non-parametric and can vary with radius. In \citetalias{Breddels2013a}, the authors find that the inner slope of the DM density profile remains unconstrained between 0 and 1.2, excluding very steep profiles $\gamma > 1.5$. However, they still find the maximum of their PPD at $\gamma = 0$, as in our inference. %However, it is not enough to exclude cuspy profiles. 
In \citetalias{Breddels2013b} the authors find that, according to $\chi ^2$ values, a cusp is preferred. However, the evidence is "barely worth mentioning" according to the Jeffrey's scale. In contrast with our results and with the measurements of \cite{Massari2018}, they found that the overall population of Sculptor shows tangential orbits ($\beta < 0$). 

\citetalias{Richardson2014}: In this work, the authors perform spherically symmetric Jeans modeling using the virial shape parameter, which provide additional constraints on anisotropy through higher-order virial equations. They use the data from \cite{Walker2009a} and bin them using circular radius. They assume that all the stars belong to a single population and model the stellar component with a Plummer model. For the DM, they explore two different approaches. First, they fix the profile to a NFW profile (cusped) and a Burkert profile (cored) \citep{Burkert1995}. Second, they use a Zhao profile \citep{Zhao1996} with a free inner slope. The velocity anisotropy profile can vary freely with the radius. For the fixed DM density profile case, they found that, according to $\chi ^2$, a cored DM profile fits the data better than the NFW one. However, when they let the inner and outer slopes free, there is a lot of degeneracy. Both a cusp with an outer slope > 2.5 or a core with an outer slope of around 2 fit the data equally well. This could hint that comparing two DM density profiles according to the evidence of two different models, could be leading to false constrained results, and once should instead use a more general model. In contrast with this work we find a core with an outer slope larger than 2.5, but the differences could arise from the different datasets, models and method used.

\citetalias{Zhu2016}: In this work, the authors fit the data using axisymmetric Jeans modeling and including rotation in their models. They fit individual stars from the dataset of \cite{Walker2009a}. The study fits the two populations independently, using the metallicity to distinguish populations. The stellar density profiles are computed assuming a Gaussian expansion, and the stars are considered tracers. For the DM density profile, they use a generalized NFW distribution with a free inner slope. The velocity distribution is approximated as Gaussian with added rotation and a radially constant anisotropy. They find an inner slope of $0.5 \pm 0.3$ (1-$\sigma$). The PPD is slightly bimodal, with peaks at around 0.25 and 0.75, but in general the result is consistent with ours.

\citetalias{Strigari2017}: This study uses dynamical modeling with energy and angular momentum DFs for the stars. They use the data from \cite{Battaglia2008} binned in semi-major axis radius and \cite{Walker2009a} binned in circular radius, independently. They use two different potentials, a NFW profile (cusped) and a Burkert profile (cored). Two populations are fitted simultaneously, with a free anisotropy parameter allowing for a transition from radial to tangential orbits both at the center and at larger distances. The $\chi^2$ comparison shows a slightly better fit for a cored profile when analyzing the data from \cite{Battaglia2008}. However, when analyzing the data from \cite{Walker2009a} the preferred model is cusped.

\citetalias{Kaplinghat2019}: In this work, the authors perform Jeans modeling. They use binned data from \cite{Walker2009a}, assuming only one population. They assume a Plummer profile for the stellar component. For the DM component, they try two different models, a NFW and a cored isothermal profile \citep{Kaplinghat2016}. The velocity anisotropy can vary with the radius, allowing both radial and tangential orbits. They find moderate to strong evidence of the cored models being preferred according to the scale of \cite{Wright2015}.

\citetalias{Read2019}: Using the software GravSphere, the authors solve the Jeans equations by fitting binned data with two virial shape parameters. They use binned data from \cite{Walker2009a}. The stellar density is modeled using a sum of three Plummer spheres. The DM distribution follows a cored NFW profile with flexible $\gamma$. Anisotropy is free, and stars are treated as tracers. The authors find a logarithmic inner slope of $\gamma = 0.8 ^{+0.3}_{-0.25}$, marginally consistent with our inference. They also find a value of $\rho_{150}=1.49^{+0.28}_{-0.23}\times 10^8 M_{\odot}/\text{kpc}^3$, while we infer a lower density $\rho_{150}=1.02^{+0.12}_{-0.13}\times 10^8 M_{\odot}/\text{kpc}^3$. Key differences between both works are: we use discrete rather than binned data, we are using both populations as independent tracers of the same potential, and we do a DF-based modeling while they do Jeans modeling.

\citetalias{Hayashi2020}: In this work, the authors apply axisymmetric Jeans modeling. They use the data from \cite{Walker2009a}, fitting individual stars and assuming only one population. For the stellar component, they fit an axisymmetric Plummer profile and assume that the stars are tracers. For the DM density, they use a double power-law density profile. They find a value of the logarithmic inner slope of $\gamma = 0.45 ^{+0.41}_{-0.31}$, which is within 1-$\sigma$ with our results. 

\citetalias{exposito-marquez2023}: The authors use a probabilistic deep learning approach, trained on velocity maps of pressure supported galaxies from NIHAO simulations. This technique applied is applied to \cite{Walker2009a} data, and it predicts a cusp for Sculptor. This suggests that cusped galaxies in these simulations resemble the observed characteristics of Sculptor better than those with cored profiles. However, it remains to be proven if these simulations accurately represent observed galaxies in this mass regime.

\section{Discussion}\label{sec:Discussion}
In this section, we discuss the main results of this work. In Sec.~\ref{sec:Dis_core}, we review the implications of Sculptor having a DM halo density profile deviating from a pure-NFW in the context of feebdback-induced effects. In Sec.~\ref{sec:Dis_JD}, we analyze how suitable Sculptor is as a candidate target for DM detection.

\subsection{Sculptor's DM halo properties accounting for stellar feedback}\label{sec:Dis_core}

According to the studies in the literature with the most efficient transformations of cusps into core due to supernovae feedback, one would expect cores to be formed around $M_{\star}/M_{halo} = 10^{-2}$ \citep{Dicintio2014, onorbe2015} and cusps being preserved around $M_{\star}/M_{halo} = 10^{-4}$. Assuming that LG dwarf galaxies follow the mass-dependent density profile proposed in \cite{DiCintio2014b}, \cite{Brook2015} estimate $\gamma \sim 0.6$, in line with our determination. The possibility that Sculptor's DM halo density profile has been affected by supernovae feedback is also aligned with the work by  \cite{Bermejo-Climent2018}, who, on the basis of the estimated amount of energy deposited by supernovae feedback based on star formation histories from deep CMD predict that Sculptor and Fornax are the MW dSphs most likely to host cored DM haloes.

Nonetheless, there is still significant uncertainty in the predictions. For example, some higher resolution simulations show that approximately constant density cores with core sizes of the order of the stellar half-light radius \citep[e.g.][]{Read2016} or up to 0.5-1kpc \citep{onorbe2015} can form even for $M_\star/M_{halo}\sim 10^{-4}$; although, a condition for this to happen is substantial late-time star formation. In particular, recently, \cite{Muni2025} have found the central DM density (quantified through $\rho_{\rm 150}$) to correlate strongly with (and to be a decreasing function of) the ratio between the stellar mass formed post- and before reionization. Observationally, it is hard to quantify the amount of stellar mass formed before and after re-ionization; an handier indicator could be the times in which 50 and 90 percent of the stellar mass formed. In any case, the latest results in the literature indicate that most star formation in Sculptor occurred in the first Gyrs of evolution, with the stellar mass essentially in place at a lookback time of 10-11 Gyr \citep[e.g.][]{Bettinelli2019} or extending out to 4-6 Gyr ago but still with the great majority of star formation occurring early on, out to 8-9 Gyr ago \citep[e.g.][]{deboer2012, Savino2018}. In addition, our dynamical analysis suggests an almost constant inner slope out to a radius about twice the half-light radius. These are all interesting constraints that simulations can aim at reproducing. 

Since, as highlighted in \cite{Muni2025}, the potential to overcome for core creation is not that of the whole halo but rather that of the concentrated central regions, in Tab.~\ref{tab:extra-parameters} we also provide the DM mass within the half-light radius and 2 kpc. 

\subsection{Sculptor as a target for indirect DM searches}\label{sec:Dis_JD}

Dwarf galaxies are among the best targets to look for high energy emission due to DM particles annihilation or decay \citep{Albert2020, Hu2024}. This is because they are very DM dominated, close enough and with a size in the sky so that they can be almost fully observed with the aperture of the telescopes usually used for this task. Once the DM density profile of a dSph is known, it is straightforward to predict the gamma flux due to emission from annihilation or decay of DM particles as a function of the aperture observed. The $\gamma$-ray flux is proportional to the so-called J (annihilation) and D (decay) factors. They can be computed as:

\begin{equation}
    J(\theta)=\frac{2 \pi}{d^2} \int_{-\infty}^{+\infty} \mathrm{d} z \int_0^{\theta d} \rho_{\mathrm{dm}}^2 R \mathrm{~d} R
\end{equation}
and 
\begin{equation}
    D(\theta)=\frac{2 \pi}{d^2} \int_{-\infty}^{+\infty} \mathrm{d} z \int_0^{\theta d} \rho_{\mathrm{dm}} R \mathrm{~d} R,
\end{equation}

\noindent where $\theta = R/d$ is the angular distance from the center of the galaxy, $z$ indicates the l.o.s. direction, $d$ is the distance to the galaxy, and $\rho_{DM}$ is the DM density. In Tab.~\ref{tab:extra-parameters} we provide the J- and D- factors computed for $\theta=0.5^{\circ}$, for the case of semi-major axis radius. The factors computed by \citetalias{Hayashi2020} are larger than ours, but within 2-$\sigma$, because of the denser profile at the center. Interestingly, compared with \cite{Geringer-Sameth2015}, we find a lower value for $J(0.5^\circ)$ but a considerably larger one for $D(0.5^\circ)$, this is explained by different shapes in the DM density profile. Very similar results were already quoted with these models in \citep{Nipoti2024}, the very minor differences are because of the use of a slightly different photometric center for Sculptor.

\section{Conclusions}\label{sec:conclusions}

In this work, we used dynamical models based on distribution functions (DFs) depending on actions to fit state-of-the-art measurements of the kinematics and metallicities of Sculptor member stars, aiming to infer the DM density distribution of the galaxy. In these models we fit discrete data in the form of individual l.o.s. velocities, metallicities and positions. We rely on a full Bayesian framework to constrain the PPD of all models parameters. We identified the two main stellar populations of this galaxy, with a median metallicity around [Fe/H]$=-1.4$ and $-2.0$, and confirmed the presence of a third population, in agreement with previous works; probabilities of membership were computed based on these models. In the following, we report our main results:

\begin{itemize}
\item We find a ratio of dark-to-stellar mass of $6^{+4}_{-3}$, $13.3^{+2.3}_{-1.2}$ and $154^{+46}_{-32}$ at 0.03 kpc, 0.32 kpc ($r_h$) and 2 kpc, with 2 kpc being our outermost measured point. The latter value sets an upper limit of $M_{\star}/M_{halo}$ around 5-8$\times10^{-3}$, providing a constraint for abundance matching relations on small galactic scales and for models of stellar feedback induced transformations of initially cuspy DM haloes.
    \item We find that the DM density profile in the inner region of the galaxy has a inner logarithmic slope of $\gamma = 0.39^{+0.23}_{-0.26}$ and a PPD shifted towards 0. This slope extends to $\rs=0.79^{+0.38}_{-0.17}$ kpc, a value that exceeds the half-light radius of the stellar component $r_h$. The DM density profile also exhibits a rather steep transition between the inner and outer slopes. Such a profile in a faint dwarf galaxy with a short SFH like Sculptor can place constraints in DM models or feedback implementations in simulations. 
    \item We confirmed that the third stellar component found by \citetalias{ArroyoPolonio2024} did not arise due to the limitations of the velocity and position distributions used in that work. It still remains to be tested whether a more flexible metallicity distribution that considers the inclusion of tails would remove the existence of this population, but still reproduce its velocity distribution shifted from Sculptor's systemic velocity. 
    \item We tested the dynamical modeling technique presented here on two mock galaxies, one with a cuspy DM halo and one with a DM halo exhibiting a constant density core of size similar to Sculptor half-light radius. These mocks resemble the characteristics of Sculptor stellar populations and the spectroscopic data-set being analyzed. We find that the multi-population fitting is able to retrieve the presence of a cusp within 1-$\sigma$, hence if a cusp were present we should have been able to detect it. Similarly, the cored DM density profile is also well recovered by our methodology. Interestingly, we find that when treating the stellar component as a single population we are not able to recover the cusped DM distribution within 1-$\sigma$ and barely within 3-$\sigma$. 
    \item We also analyzed the effect of adopting different choices for labeling the radius of the stars (semi-major axis radius, circularized and circular) to compare with the radius in the spherically symmetric models. We found that results are consistent with each other. Moreover, in all the cases, the logarithmic slope of the DM halo density profile appears to be incompatible with a NFW-profile at a 3-$\sigma$ confidence level within the radial range below the half-light radius of the stellar component.
    \item  Our models predict a velocity anisotropy parameter profile that starts at $\beta$ = 0 in the center and gradually becomes radially biased for both the MR and MP components. Additionally, we predict that the projected radial and tangential velocity dispersions on the plane of the sky should increase and decrease with radius, respectively, in both populations. 
    
\end{itemize}

In order to advance in our interpretation of DM density profiles of dwarf galaxies in the context of the nature of DM vs the impact of baryonic effects, a systematic study of a sample of dwarf galaxies with high-quality data and probing a range of stellar mass and star formation histories would offer the precious opportunity to accurately characterize the relationships between DM density profiles and the properties of their stellar populations and provide constraints to the still uncertain model predictions. Ideally, one would like to extend the analysis to as large radii as possible to alleviate uncertainties in the stellar-to-halo mass relation; wide-field of view multi-object spectrographs like WEAVE \citep{WEAVE}, 4MOST \citep{Skuladottir2023}, PFS \citep{PSF} will provide reliable l.o.s. velocities and metallicities for more and new members in the central regions and outskirts of Milky Way dSphs in a very efficient manner. 

An higher precision in the determination of the DM halo slope at different radii, and as close to the center as possible, is something to be sought in order to exclude more of the parameter space of possible DM halo models. The inclusion of projected tangential velocities of individual stars from upcoming proper motions measurements, like the ones already presented for some dwarfs like Draco from HST measurements \citep{Vitral2024}, and upcoming ones from GAIA DR5 and the Nancy Roman/WFIRST telescope \citep{WFIRST}, will offer a valuable opportunity in this respect, in particular if available for of the order of 1000 stars per galaxy \citep[e.g.][]{Read2021}. The methodology presented here has already been adapted to incorporate proper motions of stars without a significant increase in computational time \citep[e.g.][]{dellacroce2024}.

\begin{acknowledgements}
      J. M. Arroyo acknowledges support from the Agencia Estatal de Investigación del Ministerio de Ciencia en Innovación (AEI-MICIN) and the European Social Fund (ESF+) under grant PRE2021-100638.

J. M. Arroyo, G. Battaglia, G. Thomas  acknowledge support from the Agencia Estatal de Investigación del Ministerio de Ciencia, Innovación y Universidades (MCIU/AEI) under grant "EN LA FRONTERA DE LA ARQUEOLOGÍA GALÁCTICA: EVOLUCIÓN DE LA MATERIA LUMINOSA Y OSCURA DE LA VÍA LÁCTEA Y LAS GALAXIAS ENANAS DEL GRUPO LOCAL EN LA ERA DE GAIA. (FOGALERA)", the European Regional Development Fund (ERDF) with reference PID2023-150319NB-C21 and PID2020-118778GB-I00 and the AEI under grant number CEX2019-000920-S.
E. Vasiliev acknowledges support from an STFC Ernest Rutherford fellowship (ST/X004066/1) and from the Severo Ochoa Centres of Excellence programme (CEX2019-000920-S).

This paper is supported by the Italian Research Center
on High Performance Computing Big Data and Quantum Computing (ICSC), project funded by European Union - NextGenerationEU - and National Recovery and Resilience Plan (NRRP) - Mission 4 Component 2 within the activities of Spoke 3 (Astrophysics and Cosmos Observations). 

The authors acknowledge the referee for the constructive report, which enhanced the quality of the manuscript.
\end{acknowledgements}


\bibliographystyle{aa}
\bibliography{ref.bib}

\begin{appendix}

\section{Monte Carlo Markov Chain} \label{app:met_MCMC}

We run our model with three populations (25 free parameters) using semi-major axis radius, circularized radius and circular radius. By doing so, we characterize the variations on the DM properties produced by this choice. We use a Monte Carlo Marcov Chain method and rely on the EMCEE package \citep{foreman-mackey2013}, a python implementation of the Affine Invariant Markov Chain Monte Carlo (MCMC) ensemble sampler \citep{Goodman2010}. For these runs we use 100 initial walkers. We also analyzed the galaxy using only the MR and MP populations, see App.~\ref{sec:individualpop}. In these cases, we have 10 free parameters, 5 for the stellar component and 5 for the DM density. For those cases, we used 60 walkers. Finally, we analyzed a cusped mock galaxy. In this case, we considered only a 2 populations model (20 free parameters) and a single population one (10 free parameters). We used 80 and 40 walkers, respectively. In all the cases the runs were long enough to obtain at least, 1000 steps after the burn-in for the 90 walkers, which we checked and ensures stability in the inference of the parameters. The specifications of all the runs can be seen in Tab.~\ref{tab:All-runs}. The median is computed as the 50$^{th}$ percentile of the PPD and the 1-$\sigma$ and 3-$\sigma$ confidence intervals are defined by the [16$^{th}$, 84$^{th}$] and the [0.15$^{th}$ and the 99.85$^{th}$] percentiles.

\begin{table}[]
    \centering
    \title{EMCEE runs specifications}
    \begin{tabular}{|c|c|c|c|c|c|}
    \hline
         Model & Prms. & Wlks. & Steps & Burn. & Thin. \\
         \hline
         3 pop ($R$) & 25 & 90 & 4000 & 1000 & 24 \\
         3 pop ($R_{cz}$) & 25 & 90 & 2725 & 1000 & 24 \\
         3 pop ($R_{c}$) & 25 & 92 & 1856 & 800 & 24 \\
         Only MP & 10 & 50 & 2000 & 600 & 24\\
         Only MR & 10 & 50 & 2000 & 600 & 24\\
         Cuspy mock 2-pop & 20 & 80 & 4989 & 2000 & 24 \\
         Cuspy mock 1-pop & 10 & 40 & 3277 & 1500 & 24\\
         Cored mock & 20 & 80 & 5000 & 3000 & 24\\
         \hline
    \end{tabular}
    \caption{Here we show all the specifications for the EMCEE runs. Col.~1 lists the models; Col.~2 the number of free parameters; Col.~3 indicates the final number of walkers used ; Col.~4 lists number of steps; Col.~5 the number of burning steps; and finally, Col.~6 lists the value of the thinning. We have checked that there are enough steps after the burning to ensure convergence and that the value of the thinning do not have a strong impact on the inference.}
    \label{tab:All-runs}
\end{table}

\section{Bayesian inference} \label{app:cornerpolots}
\begin{figure*}
    \sidecaption
        \includegraphics[width=1.35\columnwidth]{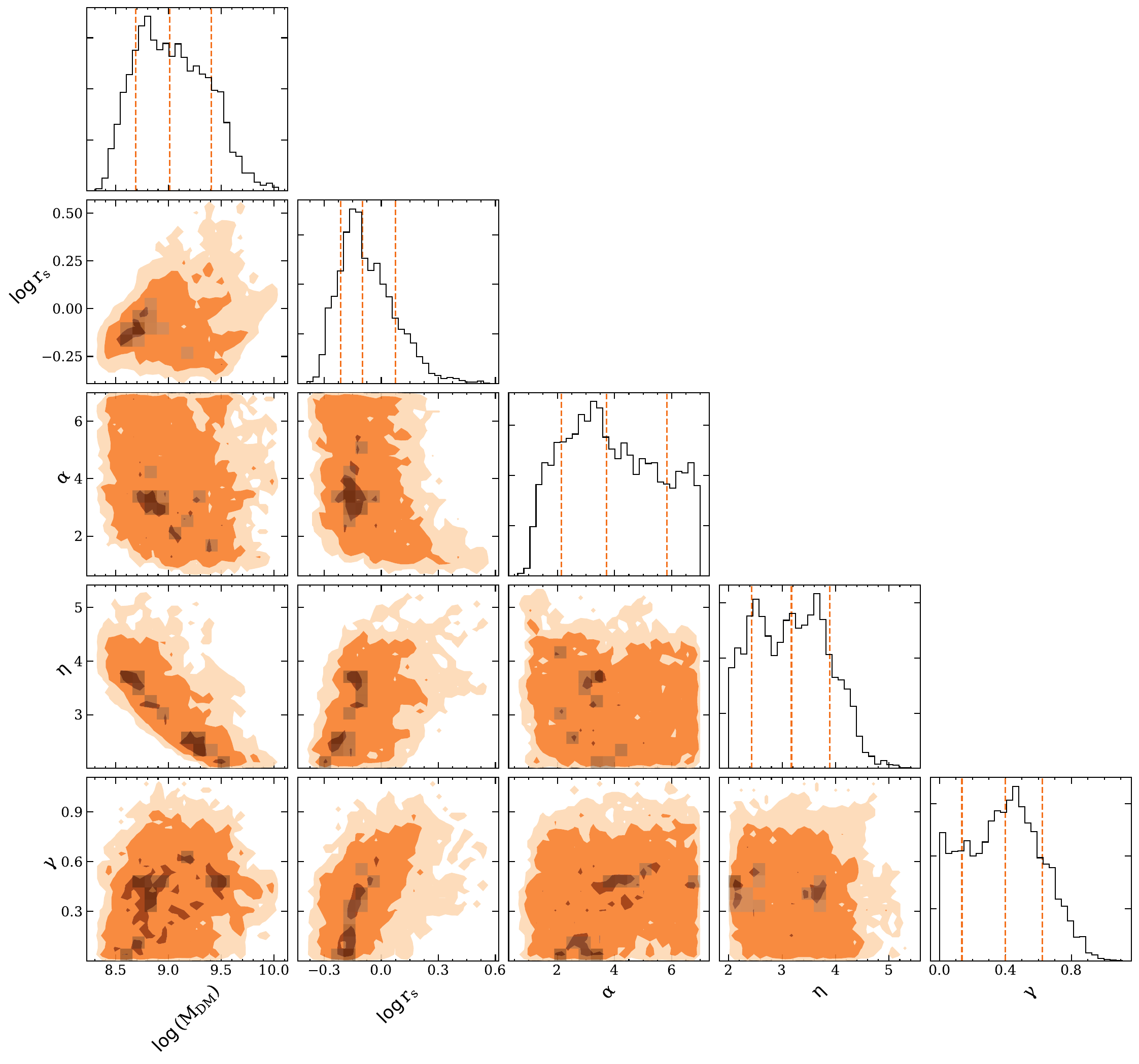}
        \caption{Corner-plot showing the PPD for the free parameters of the DM halo. The first parameter is the total mass $M_{DM}$ in $M_\odot$, the second the scale-radius $r_s$ in kpc and the third one the steepness of the transition $\alpha$ between the outer slope  (fourth parameter, $\eta$) and the inner slope (fifth parameter, $\gamma$).}\label{Fig:cornerplot_DM}
        
    \end{figure*}

\begin{figure*}
    \sidecaption
        \includegraphics[width=1.35\columnwidth]{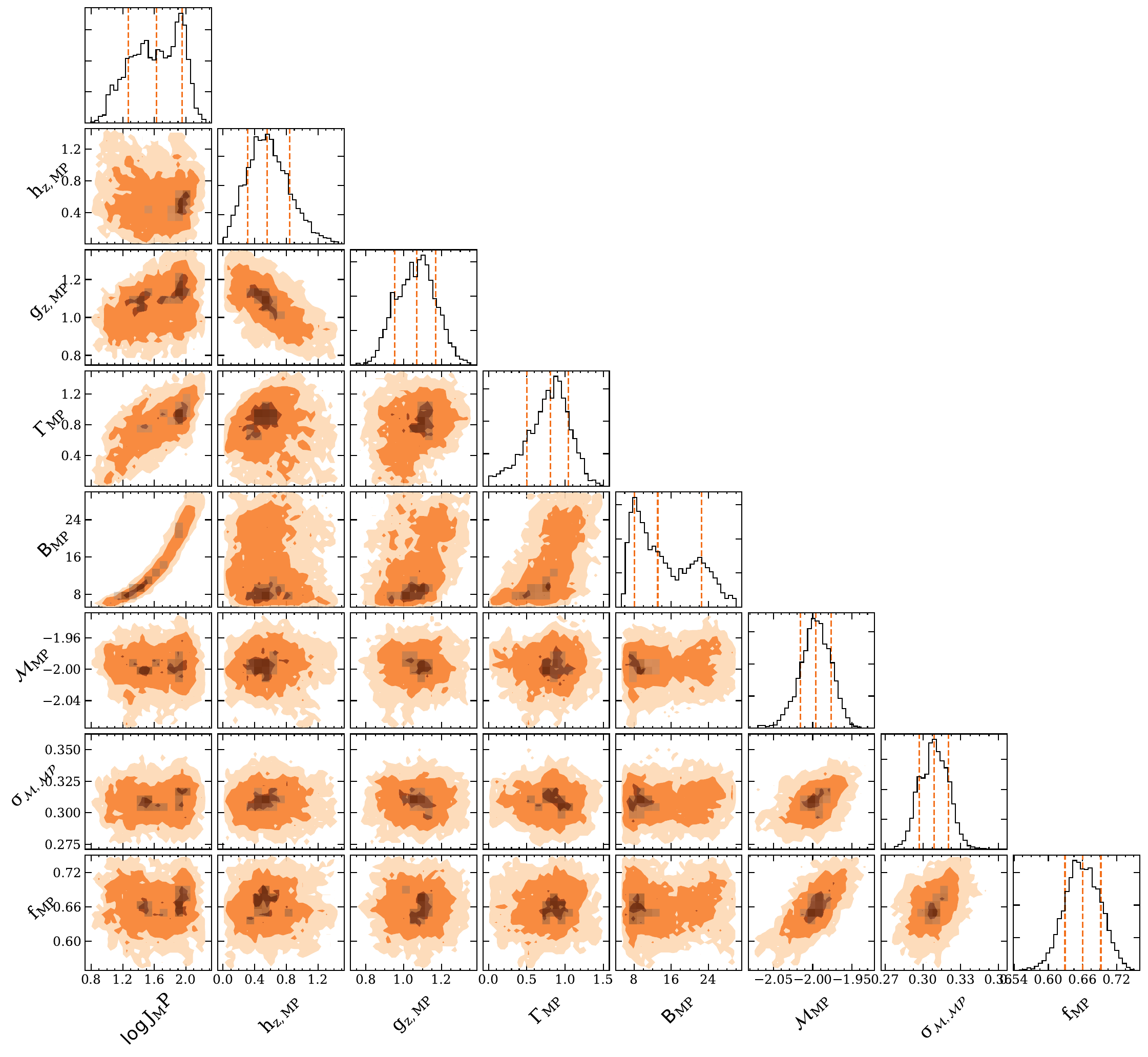}
        \caption{Corner-plot showing PPD of the free parameters of the MP component. The eight parameters shown \{$\JMP, \GammaMP, \BMP, \gMP, \hMP,$ $M_{\MP}$, $\sigma_{\mathcal{M},MP}$, $f_{MP}$\} are described in Sec.~\ref{subsec:met_sd}}
        \label{Fig:Cornerplot_MP}
\end{figure*}

\begin{figure*}
    \sidecaption
        \includegraphics[width=1.35\columnwidth]{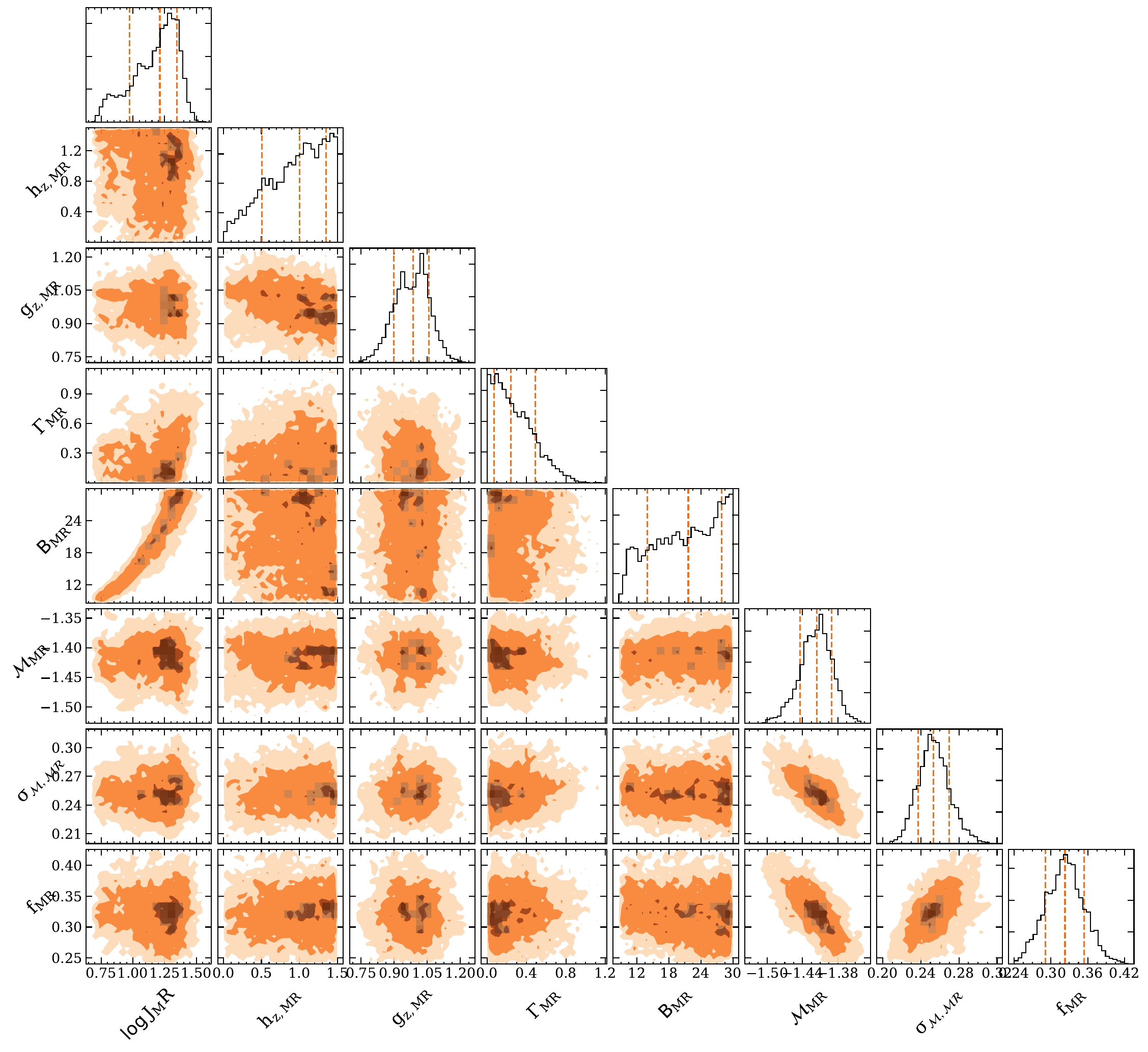}
        \caption{Corner-plot showing PPD of the free parameters of the MR component. The eight parameters shown \{$\JMR, \GammaMR, \BMR, \gMR, \hMR,$ $M_{\MR}$, $\sigma_{\mathcal{M},MR}$, $f_{MR}$\} are described in Sec.~\ref{subsec:met_sd}}
        \label{Fig:Cornerplot_MR}
\end{figure*}

\begin{figure*}
    \sidecaption
        \includegraphics[width=1.35\columnwidth]{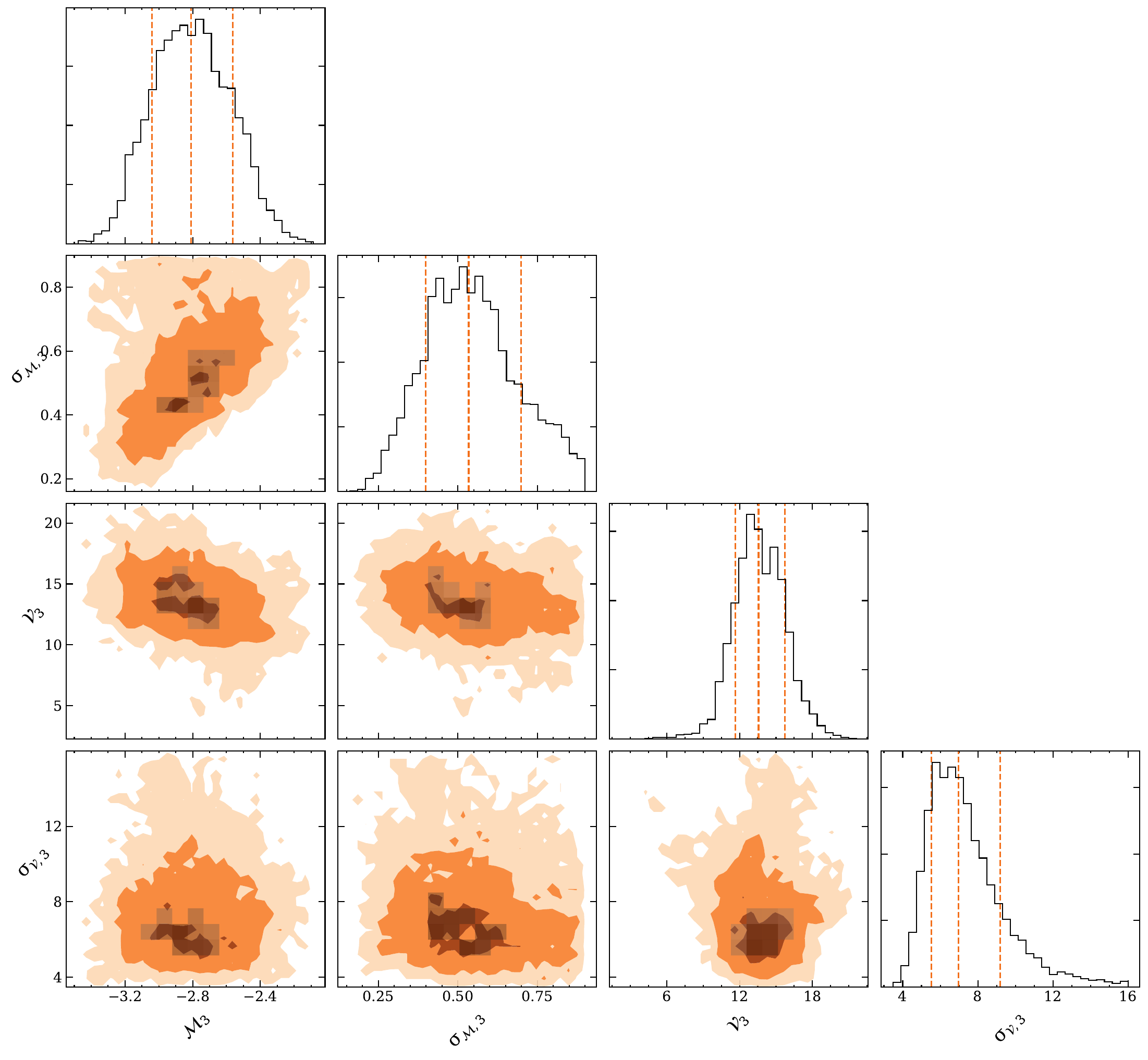}
        \caption{Corner-plot showing PPD of the free parameters of the third stellar component. The four parameters shown \{$\mathcal{V}_3$, $\sigma_{\mathcal{V},3}$, $\mathcal{M}_3$, $\sigma_{\mathcal{M},3}$\} are described in Sec.~\ref{subsec:met_sd}}
        \label{Fig:Cornerplot_POP3}
\end{figure*}

In this section, we show the inference of all the 25 free parameters for the semi-major axis radius case split in different groups. In Fig.~\ref{Fig:cornerplot_DM} we show the inference for the DM, in Fig.~\ref{Fig:Cornerplot_MP} for the MP component, in Fig.~\ref{Fig:Cornerplot_MR} for the MR component and in Fig.~\ref{Fig:Cornerplot_POP3} for the third population.


\section{Inference with individual populations}\label{sec:individualpop}

\begin{figure}
    \centering
        \includegraphics[width=1\columnwidth]{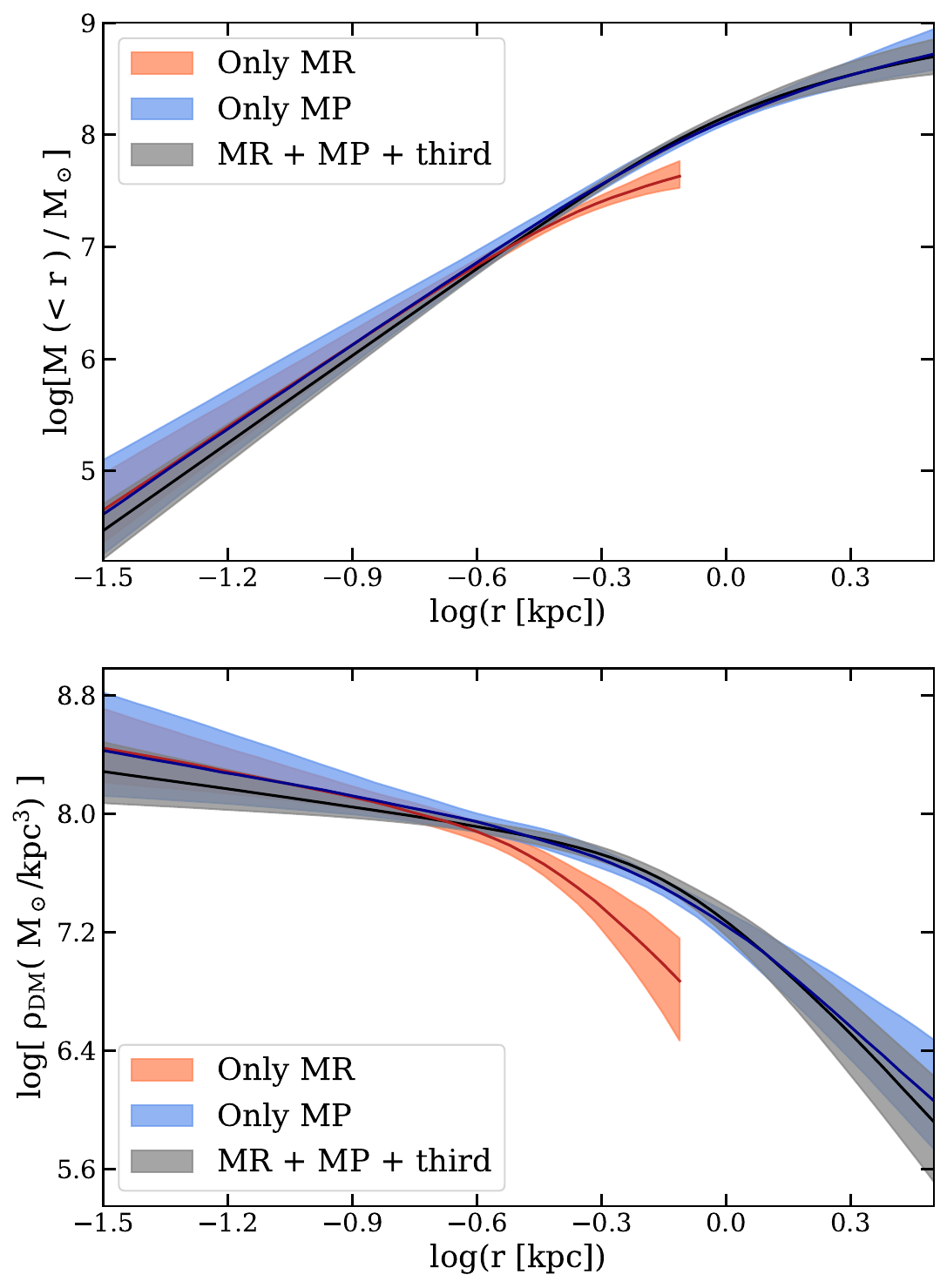}

        \caption{Upper panel: Cumulative DM halo mass distribution. Lower panel: DM halo density distribution. In different colors we show the inferred DM density profile obtained by using the two independent populations as tracers (black), only the MR population (red) and only the MP population (blue), as indicated in the legend.}
        \label{fig:Cumulat_den_profiles_indiv}
    \end{figure}

In this section, we compute the DM density profile of Sculptor using only the MR and MP populations independently. Assign membership to each star to the population it has the highest probability to belong, accourding to the probabilities of membership from \citetalias{ArroyoPolonio2024}. In Fig.\ref{fig:Cumulat_den_profiles_indiv} we show the results of the analysis. The results are consistent across all the range except in the outer part for the MR component. This could be explained by the lack of MR stars in that region. The main difference is that the profiles are considerably less constrained, and they have larger errors. As a result, the inference on $\gamma$ of individual populations fitting is in marginal agreement with both, cusped and cored profiles. For only the MR population the median of $\gamma$ is 0.50, while the 1- and 3-$\sigma$ intervals are [0.19, 0.91] and [0.00, 1.51]. While for only the MP population the median of $\gamma$ is 0.45, while the 1- and 3-$\sigma$ intervals are [0.12, 0.94] and [0.00, 1.34] This highlights the importance of having two independent tracers of the potential and the statistics necessary to constrain the inner slopes.
  
\section{Test on a Sculptor-like mock with a cored DM halo}\label{app:mock}
\begin{figure}
        \includegraphics[width=1\columnwidth]{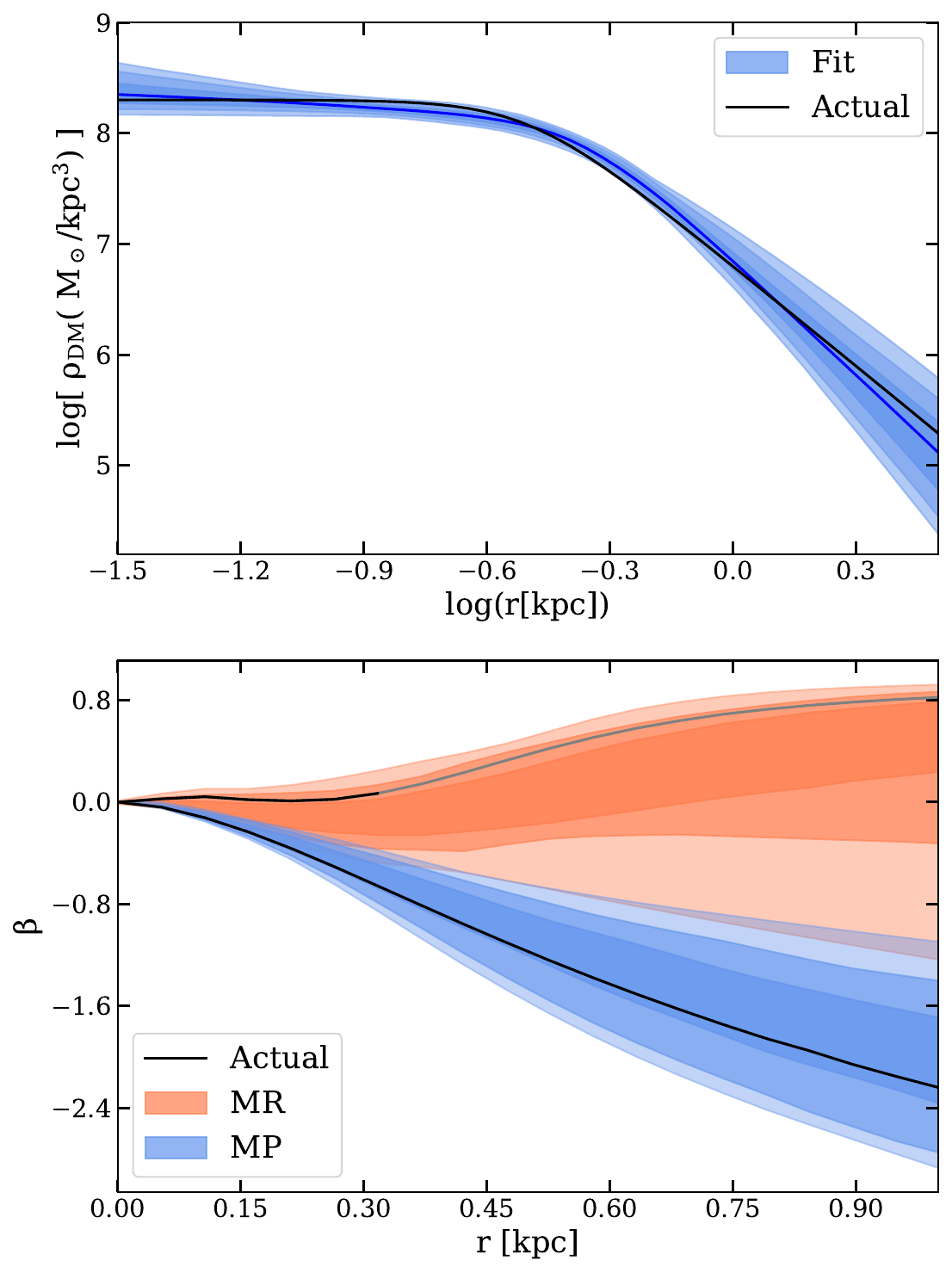}
        \caption{Upper panel: DM density profile recovered for the cored mock galaxy. In black, we show the underlying one, the median of the predicted profile is shown with a solid blue line and the blue bands indicate the 1-, 2- and 3-$\sigma$ confidence intervals. Lower panel: Velocity anisotropy profile for the MP and MR stellar components. As solid lines, we show the values for the underlying model and in bands the best-fit for the MR and the MP component, indicating the 1-, 2- and 3-$\sigma$ confidence intervals. The solid black line corresponding to the MR component become gray in the region in which we do not produce stars that belong to that component.}
        \label{fig:Mock_2pop-core}
    \end{figure}

Here, we repeat the experiment we described on Sec.~\ref{sec:Mock}, but using a cored DM halo. The core has a scale radius of 0.32 kpc, similar to the half-light radius of the stellar component in Sculptor \citep{Muñoz2018}. A comparison between the inferred parameters and their true values is presented in Tab.~\ref{tab:Mock-test-core}, while the resulting DM density and $\beta$ profiles are shown in Fig.~\ref{fig:Mock_2pop-core}. Both the inferred parameters and the profiles are in very good agreement with the underlying ones, there are some minor tensions in $r_s$ and $\gamma$ discussed in Sec.~\ref{sec:Mock}.

\begin{table*}[h]
\centering
\caption{Inference on models free parameters for the cored mock.}
\renewcommand{\arraystretch}{1.25}
\begin{tabular}{c|lcccccc}
\hline
\hline
Component & Parameter & Prior & Median & $1\sigma$ & $3\sigma$ & Actual\\
\hline
\hline
 & $\log \JMP$ & [0.3,2.5] & 0.94 & [0.64, 1.09] & [0.31, 1.20] & 0.8\\
 & $\hzMP$ & [0, 1.5] & 0.11 & [0.01, 0.26] & [0.00, 0.89] & 0.05\\
 & $\gzMP$ & [0, 1.5] & 1.07 & [0.93, 1.19] & [0.58, 1.29] & 1.2\\
MP & $\GammaMP$ & [0, 3] & 1.04 & [0.79, 1.27] & [0.56, 1.52] & 1.0\\
 & $\BMP$ & [3, 25] & 17.72 & [11.24, 21.82] & [7.22, 24.86] & 14\\
 & $\MM_{\MP}$ [dex] & [-2.1, -1.7] & -2.00 & [-2.01, -1.98] & [-2.03, -1.95] & -2\\
 & $\sigma_{\MM,\MP}$ [dex] & [0.15, 0.45] & 0.32 & [0.31, 0.33] & [0.29, 0.35] & 0.3\\
 & $\fMP$ & [0.5, 0.75] & 0.59 & [0.56, 0.61] & [0.53, 0.64] & 0.6\\
\hline

 & $\log \JMR$ & [0.3,2.5] & 1.52 & [1.34, 1.67] & [1.15, 1.87] & 1.5\\
 & $\hzMR$ & [0, 1.5] & 0.38 & [0.13, 0.86] & [0.00, 1.48] & 0.6\\
 & $\gzMR$ & [0, 1.5] & 0.45 & [0.38, 0.51] & [0.28, 0.72] & 0.4\\
MR & $\GammaMR$ & [0, 3] & 0.39 & [0.11, 0.74] & [0.01, 1.23] & 0.4\\
 & $\BMR$ & [3, 25] & 18.78 & [13.74, 23.31] & [10.34, 24.85] & 20\\
 & $\MM_{\MR}$ [dex] & [-1.65, -1.3] & -1.42 & [-1.44, -1.40] & [-1.47, -1.37] & -1.4\\
 & $\sigma_{\MM,\MR}$ [dex] & [0.15, 0.45] & 0.32 & [0.31, 0.33] & [0.28, 0.36] & 0.3\\
 & $\fMR$ & [0.25, 0.5] & 0.41 & [0.39, 0.44] & [0.36, 0.47] & 0.4\\
\hline

 & $\log \MDM$ [$M_{\odot}$] & [7, 11] & 8.41 & [8.26, 8.61] & [8.18, 8.92] & 8.5\\
 & $\log \rs$ [kpc] & [-3, 1] & -0.36 & [-0.43, -0.29] & [-0.51, -0.18] & -0.5\\
DM & $\alpha$ & [0, 7] & 4.65 & [2.82, 6.37] & [1.51, 6.97] & 4.0\\
 & $\eta$ & [2, 7] & 3.49 & [3.00, 4.17] & [2.58, 4.82] & 3.0\\
 & $\gamma$ & [0, 1.9] & 0.17 & [0.04, 0.34] & [0.00, 0.56] & 0.0\\
\hline
\hline
\end{tabular}
\tablefoot{The table is divided in three parts, they show the parameters of the MP, MR stellar populations and the DM component. Col.~1 indicates the components the parameters are referred to; Col.~2 lists the parameters; Col.~3 shows the range of uniform priors we used in the EMCEE runs; Cols.~4,~5 and~6 lists the median values, $1\sigma$ and $3\sigma$ confidence intervals; and finally Col.~7 shows the real value used to produce the mock. Details on the MCMC procedure and on how the 1- and 3-$\sigma$ confidence intervals are computed are given in App.~\ref{app:met_MCMC}}\label{tab:Mock-test-core}
\end{table*}

\end{appendix}
\label{LastPage}
\end{document}